\documentclass[12pt, draftclsnofoot, onecolumn]{IEEEtran}
\usepackage{times}
\usepackage{epsfig}
\usepackage{graphicx}
\usepackage{textcomp}
\usepackage{amsmath}
\usepackage{amsfonts}
\usepackage{cite}
\usepackage{psfrag}
\usepackage{subfigure}
\usepackage{stfloats}
\usepackage{amsmath}
\usepackage{amssymb}
\usepackage{array}
\usepackage[noend]{algorithmic}
\usepackage{algorithm}
\usepackage{multicol}
\usepackage{multirow}
\usepackage{color}
\usepackage{bm}
\usepackage{setspace}
\usepackage{float}
\usepackage{soul}
\usepackage{color}

\newtheorem{proposition}{\textbf{Proposition}}
\newtheorem{corollary}{\textbf{Corollary}}
\newtheorem{lemma}{\textbf{Lemma}}

\begin{document}
\title{\LARGE{Joint Power Allocation and Decoding Order Selection for NOMA Systems: Outage-Optimal Strategies}}
\author{Mengqi~Yang,~Jian~Chen,~Zhiguo~Ding, Yuanwei Liu, Lu Lv, and Long Yang
\thanks{Mengqi Yang, Jian Chen, Lu Lv, and Long Yang are with the State Key Laboratory of Integrated Services Networks, Xidian University, Xi'an 710071, China (e-mail: mqyang\_xidian@hotmail.com; jianchen@mail.xidian.edu.cn; \{lulv, lyang\}@xidian.edu.cn).
Zhiguo Ding is with the School of Electrical and Electronic Engineering, the University of Manchester, Manchester M13 9PL, U.K. (e-mail: zhiguo.ding@manchester.ac.uk).
Yuanwei Liu is with the School of Electronic Engineering and Computer Science, Queen Mary University of London, London E1 4NS, U.K. (e-mail: yuanwei.liu@qmul.ac.uk). This work has been submitted to the IEEE for possible publication. Copyright may be transferred without notice, after which this version may no longer be accessible.
}\vspace{-15mm}
}
\maketitle
\pagestyle{plain}

\begin{abstract}\vspace{-3mm}

  We investigate joint power allocation and decoding order selection (PA-DOS) aimed at enhancing the outage performance of non-orthogonal multiple access (NOMA) systems. By considering the diverse target rates of users, new important properties of NOMA are revealed: When users' target rates satisfy certain conditions, the channel state information (CSI) is not required by PA-DOS to minimize the system outage probability, and different users' outage probabilities can be minimized simultaneously; When such conditions are not satisfied, the opposite situation occurs. Following these properties, two PA-DOS strategies are designed regarding distinct user priorities, which ensure the minimization of the system outage probability and the user outage probability of the high-priority user. Especially, these strategies do not require CSI or only require one-bit CSI feedback depending on users' target rates. Analytical and numerical results are provided to demonstrate that the proposed strategies significantly outperform the existing strategies in terms of both system and user outage performance. Furthermore, the results show that although under some target rates the minimum user outage probabilities cannot be simultaneously achieved, they can be closely approached at the same time in the high-signal-to-noise-ratio regime by the proposed strategies.
\end{abstract}\vspace{-3mm}
\begin{IEEEkeywords}\vspace{-3mm}
Decoding order, non-orthogonal multiple access (NOMA), outage probability, power allocation.
\end{IEEEkeywords}\vspace{-3mm}
%
%
\section{Introduction}
As a promising candidate multiple access technique in 6G mobile networks, non-orthogonal multiple access (NOMA) shows its advantages on providing more access opportunities and improving spectral efficiency compared to the orthogonal multiple access (OMA) \cite{Yuan_survey,Islam_survey}. In NOMA, the transmitter allocates its power for multiple data flows with a set of power coefficients, and then uses superposition coding to impose data flows in one signal \cite{NZhao_TCOM,YXu_adaptive}. At the receiver, successive interference cancellation (SIC) is employed to decode data flows in a sequential manner with a selected decoding order \cite{YXu_robust,Ycao_TWC,HLei_TCOM}. As suggested by this technical principle, power allocation and decoding order selection (PA-DOS) are crucial and interrelated issues to NOMA, which largely affect the overall system performance.

\subsection{Research Background}
According to the adopted patterns of the power coefficients, NOMA can be classified in two types, namely NOMA with dynamic power coefficients (D-NOMA) and NOMA with fixed power coefficients (F-NOMA).
In D-NOMA, the power coefficients are dynamically adjusted according to the instantaneous channel states \cite{Ding_impact,BChen_CRNOMA,FZhou_CRNOMA,ZXiao_fairness,ZYang_general}. By this way, D-NOMA has a high flexibility on PA, which can dynamically allocate any portion of the transmit power to one user in each transmission block.
One typical example for D-NOMA is the cognitive radio inspired NOMA \cite{Ding_impact,BChen_CRNOMA,FZhou_CRNOMA}, where the high-priority user decodes its message by treating the low-priority user's data flow as interference, while PA is conducted to constrain the interference level. As the result, the quality of service of the high-priority user is ensured, and the low-priority user can be served opportunistically. For fairness consideration, PA was further studied in \cite{ZYang_general} to provide each user a data rate not lower than the counterpart in OMA, while the work in \cite{ZXiao_fairness} aimed at maximizing the minimum data rate achieved by all the users.

On the other hand, in F-NOMA power coefficients are fixed and predetermined before the data transmission. Despite the employment of fixed power coefficients, F-NOMA still reserves a simple flexibility on PA during the data transmission, i.e., power coefficients can be adaptively assigned to different data flows.
One most widely adopted paradigm for joint PA-DOS in F-NOMA is the channel state determined (CSD) strategy proposed in \cite{Ding_random_users}. The CSD strategy suggests that, in each transmission block the greater power coefficient should be adaptively assigned to the user with the weak instantaneous channel state, i.e., more power is allocated to the weak user. As the associated decoding order, the weak user's message will be decoded first at each user \cite{Ding_random_users}. With employing the CSD strategy, user outage probabilities were evaluated in \cite{Ding_random_users}, system outage probability was analyzed in \cite{THou_nakagami,ZYang_imperfectCSI}, and the values of power coefficients were optimized in \cite{Ionnis_fairness,GLi_RA}.

Obviously, F-NOMA underperforms D-NOMA due to the lower flexibility in PA. Nevertheless, as the advantage, F-NOMA has the potential to be implemented without full instantaneous channel state information at the transmitter (CSIT), which shows great importance in massive-device applications such as the internet of things \cite{Sha_grantfree,XLi_B5G}. To enable the implementation of F-NOMA with limited CSIT, there are several types of PA-DOS strategies developed in existing works \cite{PXu_1bit,Kim_CMD,Gong_CMD,YLiu_CoNOMA,HWang_HARQ,GX_DD,YXu_TWC,Ding_QoS,LY_QoS,Lulv_QoS}.
In \cite{PXu_1bit}, users were first partitioned into a strong user group and a weak user group based on their one-bit feedbacks. Then CSD was performed for users in different groups, while random PA-DOS was performed for users within the same group. Following the similar rationale of CSD, PA-DOS in \cite{Kim_CMD,Gong_CMD,YLiu_CoNOMA,HWang_HARQ,GX_DD,YXu_TWC} was also conducted based on ordering users with regard to the channel conditions. Nevertheless, instead of the instantaneous channel states, the mean values of channel gains were used for user ordering in \cite{Kim_CMD,Gong_CMD}, while the transmission distances were adopted in \cite{YLiu_CoNOMA,HWang_HARQ,GX_DD,YXu_TWC}, which are referred to as the channel mean determined (CMD) strategy and the distance determined (DD) strategy, respectively.
From a very different perspective, service priority determined (SPD) PA-DOS strategy was proposed in \cite{Ding_QoS,LY_QoS,Lulv_QoS}, where more power was assigned to the high-priority user, and its message was first decoded by all the users. Moreover, in the above works, the system outage probability was investigated in \cite{PXu_1bit,GX_DD,YXu_TWC,Ding_QoS,Lulv_QoS}, and the user outage probabilities were investigated in \cite{Gong_CMD,YLiu_CoNOMA,HWang_HARQ,YXu_TWC,LY_QoS,Lulv_QoS,GX_DD}, which are two often used performance metrics for F-NOMA when the users have specific target data rates, especially for the scenario without full CSIT. For clarity, the abbreviations referring to the PA-DOS strategies in this paper are summarized in Table \ref{Table_Strategies}.

\begin{table}[t!]
\small
\renewcommand\arraystretch{1}
  \caption{Abbreviations Referring to PA-DOS Strategies}\label{Table_Strategies}
  \centering
  \begin{tabular}{|c|c|}
    \hline
      \textbf{Abbreviation} & \textbf{Phrase} \\
    \hline
      CSD  & Channel state determined \\
      CMD  & Channel mean determined \\
      DD  & Distance determined \\
      HUF  & High-rate user first \\
      LUF  & Low-rate user first \\
      SPD  & Service priority determined \\
    \hline
  \end{tabular}\vspace{-3mm}
\end{table}

\subsection{Motivations and Contributions}
For a long time, the above mentioned strategies, especially the CSD strategy, were adopted as mature solutions to PA-DOS in most existing works. However, a more recent work \cite{KSAli} showed a meaningful phenomenon that, the CSD strategy could achieve a worse outage performance than the DD strategy under some system parameters. This is contrary to the common sense, since CSD adapts PA-DOS to the exact channel states, while DD performs static PA-DOS solely depending on the path loss. As this phenomenon was only revealed by simulations in \cite{KSAli}, the detailed reasons behind it remained unknown.

In fact, a more generalized phenomenon will be shown in Section \ref{Section_simulation} of this paper, i.e., none of the above mentioned  strategies can ensure the best outage performance for F-NOMA systems.
Through analysis we find that, the key reason behind it is the lack of considering users' target rates in the design of existing PA-DOS strategies. To be specific, the outage occurs if the channel capacity is lower than the target rate, where the channel capacity is affected by the channel state and PA-DOS. Thus, the target rate is a benchmark to determine whether a channel is strong or weak, not merely depending on the comparison among channels. Meanwhile, the target rate is also a benchmark to evaluate whether the PA-DOS, which is essentially the resource allocation, is over-allocated or under-allocated for each user in NOMA. When such a crucial factor is neglected, the effectiveness of the designed PA-DOS strategy is consequently limited.
\begin{figure}
  \centering
  \includegraphics[width=2.5in]{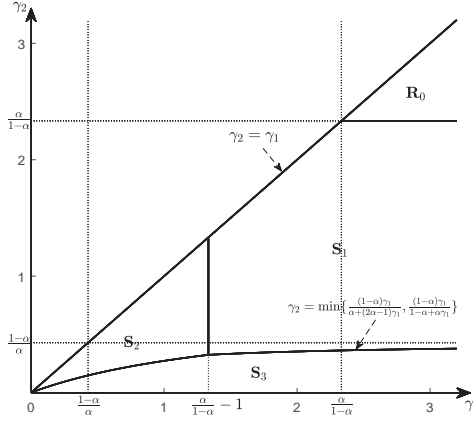}
  \caption{An illustration of $\mathbf R_0$ and $\mathbf S_k$, $k=1,2,3$, where $\mathbf R_0$ and $\mathbf S_k$ are regions of $(\gamma_1,\gamma_2)$ defined in Section \ref{Section_strategies}, $\gamma_i\triangleq 2^{R_i}-1$ for $i=1,2$, $R_1$ and $R_2$ denote users' target rates with $R_1\geq R_2$, and $\alpha$ denotes the power coefficient.}\label{fig_HUF_regions}
\end{figure}

Motivated by the above observations, we raise two fundamental questions in this paper. First, when taking the target rates into the consideration, how do they exactly affect the PA-DOS and the corresponding outage performance of F-NOMA? Moreover, since the strategies without CSIT can outperform the CSD strategy, the second question arises: Is the CSIT necessary for achieving the best outage performance, and if yes how should it be properly used?
To find the answers, we investigate the PA-DOS in a two-user downlink F-NOMA network, where both the system outage performance and the user outage performance are examined. The main contributions of this paper are summarized as follows:
\begin{itemize}
  \item Instead of following the existing strategies, the whole flexibility of PA-DOS in F-NOMA is considered, where the PA-DOS can be adjusted at each transmission block, and the DOS can be independently conducted for each user rather than determined by PA. Based on this flexibility, two rate-adaptive PA-DOS criteria are respectively proved from the perspective of the system and the users, which reveal the following important properties of F-NOMA:
      \begin{enumerate}
      \item When the target rates satisfy $(\gamma_1, \gamma_2)\in\mathbf R_0$, as illustrated by Fig. \ref{fig_HUF_regions}, the system outage and the user outage inevitably occur with arbitrary PA-DOS;
      \item When $(\gamma_1, \gamma_2)\in\mathbf S_1$, CSIT is not required by the outage-optimal PA-DOS. Specifically, simply allocating more power to the low-rate user and decoding its message first not only achieves the minimum system outage probability, but also achieves the minimum user outage probabilities for both of the two users;
      \item When $(\gamma_1, \gamma_2)\in\mathbf S_2\cup\mathbf S_3$, CSIT is crucial for PA-DOS to achieve the minimum system outage probability. Moreover, the minimum user outage probabilities cannot be simultaneously achieved for the two users by any PA-DOS strategies.
      \end{enumerate}
  \item Based on the above findings, two outage-optimal PA-DOS strategies are designed with considering different user priorities, which are termed as high-rate user first (HUF) and low-rate user first (LUF). When $(\gamma_1, \gamma_2)\in\mathbf S_1$, the proposed strategies follow the second property of F-NOMA, and thus minimize the system outage probability and the user outage probabilities without CSIT. When $(\gamma_1, \gamma_2)\in\mathbf S_2\cup\mathbf S_3$, the proposed strategies only require one-bit CSI feedback for PA-DOS, and they both achieve the minimum system outage probability. Meanwhile, the HUF (LUF) strategy achieves the minimum user outage probability for the high-rate user (low-rate user).
  \item The outage probabilities achieved by the proposed strategies are derived in closed forms. Theoretical and simulated results show that: i) the proposed strategies outperform existing strategies on both system and user outage performance. Especially, the outage error floors caused by existing strategies in some scenarios are avoided by the proposed strategies; ii) Although when $(\gamma_1, \gamma_2)\in\mathbf S_2\cup\mathbf S_3$ the minimum user outage probabilities cannot be achieved simultaneously, they can be closely approached at the same time in the high-signal-to-noise-ratio regime by either of the proposed strategies.
\end{itemize}

\section{System Model}
\begin{table}[t!]
\small
\renewcommand\arraystretch{1.1}
  \caption{Summary of Main Notations}\label{Table_notations}
  \centering
  \begin{tabular}{|c|c|c|c|}
    \hline
      \textbf{Notation} & \textbf{Description} & \textbf{Notation} & \textbf{Description}\\
    \hline
      $U_i$  & User $i$, $i=1,2$ & $\Gamma_{U_i \rightarrow x_j}^{\pi(t),\omega_i(t)}$ & SINR/SNR when $U_i$ decodes $x_j$\\
      $R_i$  & Target rate of $U_i$ & $\gamma_i$ & SINR/SNR threshold for detecting $x_i$\\
      $x_i$  & Message intended by $U_i$ & $\mathbb S^{\pi(t), \omega_i(t)}_i$ & Successful decoding event of $U_i$\\
      $t$ & Index of transmission block & $\mathcal P_{\text{sys}}$ & System outage probability \\
      $h_i(t)$ & Channel coefficient of $U_i$ & $\mathcal P_{i}$ & User outage probability\\
      $\alpha$ & Constant power coefficient & $[\pi(t),\omega_1(t),\omega_2(t)]$ & Joint decision of PA-DOS\\
      $p$ & Total transmit power of the BS & $\mathbf O(\pi(t))$ & Set of available $\omega_i(t)$\\
      $p_i(t)$ & Transmit power allocated to $U_i$ & $\mathbf P$ & Set of available $\pi(t)$\\
      $\pi(t)$ & Indicator for PA  &$\mathbf R_k$, $\mathbf S_k$, $\mathbf G_k$ & Defined regions of $(\gamma_1,\gamma_2)$\\
      $y_i^{\mathcal \pi(t)}$ & Signal observation at $U_i$ & $\phi_i^{\pi(t),\omega_i(t)}$ & Threshold for $|h_i(t)|^2$ \\
      $\omega_i(t)$ & Indicator for DOS at $U_i$  & $\Phi^{\pi(t),\omega_1(t),\omega_2(t)}$ & Function of $|h_i(t)|^2$ and $\phi_i^{\pi(t),\omega_i(t)}$\\
    \hline
  \end{tabular}\vspace{-5mm}
\end{table}

We consider a two-user downlink F-NOMA communication scenario, which consists of a base station (BS) and two users, denoted by $U_1$ and $U_2$ \footnote{
As reported by \cite{Ding_impact}, if the messages of massive users are superimposed in NOMA systems, it incurs strong co-channel interference and high computational complexity. Thus, the two-user NOMA is a simple but more realistic approach to implement NOMA in practical systems, which has been adopted by existing protocols, e.g., the multi-user superposition transmission (MUST) in 3rd-generation partnership project long-term evolution (3GPP-LTE) \cite{MUST}. Moreover, the research results under the two-user NOMA scenario can be applied to the scenario with more than two users, where user pairing is employed to construct a hybrid NOMA system, e.g., pairing users based on the distances \cite{YLiu_CoNOMA}. The research on how target rates affect the user pairing is an important direction for our future research.}.
For clarity, the main notations used in this paper are summarized in Table \ref{Table_notations}. It is assumed that, $U_1$ and $U_2$ have specific target rates for their services, denoted by $R_1$ and $R_2$, which are known by the BS. Without loss of generality, $R_1\geq R_2$ is assumed. The channels from the BS to $U_1$ and $U_2$ experience independent but non-identically distributed block fading, of which the channel coefficients are denoted by $h_1(t)$ and $h_2(t)$, respectively, with $t$ denoting the $t$-th transmission block. Thus, $|h_1(t)|^2$ and $|h_2(t)|^2$ are used to denote the channel gains.

Following the F-NOMA protocol, the BS divides its total transmit power $p$ with a constant power coefficient $\alpha$, where $1>\alpha>\frac{1}{2}$ is assumed and thus $\alpha p$ particularly denotes the part of higher power, while $(1-\alpha) p$ denoting the part of lower power. Then, in each transmission block, the BS adaptively allocates $\alpha p$ and $(1-\alpha) p$ to $U_1$ and $U_2$.
To indicate which user is allocated with the higher power at transmission block $t$, we define the indicator function $\mathcal \pi(t)$ as
\begin{align}
  \mathcal \pi(t) \triangleq
  \left\{
  \begin{array}{ll}
   1, &\text{if}~p_1(t)=\alpha p ~\text{and}~p_2(t)=(1-\alpha)p,\\
   2, &\text{if}~p_1(t)=(1-\alpha)p ~\text{and}~p_2(t)=\alpha p,
  \end{array}
  \right.
\end{align}
where $p_1$ and $p_2$ denote the amount of power allocated to $U_1$ and $U_2$, respectively. By selecting $\pi(t)$ from $\{1,2\}$ and using the superposition coding, the BS transmits a superimposed signal $x^{\pi(t)}=\sqrt{p_1(t)} x_1+ \sqrt{p_2(t)} x_2$, where $x_1$ and $x_2$ denote the messages desired by $U_1$ and $U_2$, respectively. After the transmission, the signal observation at $U_i$, $i=1,2$, can be expressed as
\begin{align}\label{signal_received}
  y_i^{\mathcal \pi(t)}= (\sqrt{p_1(t)}x_1+\sqrt{p_2(t)}x_2) h_i(t)+n_i(t),
\end{align}
where $n_i(t)$ denotes the additive Gaussian white noise with zero mean and variance being $\sigma^2$. 

To reveal the impact of target rates on F-NOMA, in this paper we consider that, each user can adaptively and independently selects its decoding order from first decoding $x_1$ and first decoding $x_2$ at each transmission block, rather than determined by PA. This flexibility indicates that, one user is allowed to first decode the message carrying the less power, which is feasible as long as the target rate of this message is very low \cite{Ding_unveiling,Common_myth}.
For simplicity, we use indicator function $\mathcal \omega_i(t)$ to express the DOS of $U_i$ at transmission block $t$, which is defined as
\begin{align}
  \mathcal \omega_i(t) \triangleq
  \left\{
  \begin{array}{ll}
   1, &\text{if~$U_i$~first~decodes~$x_1$},\\
   2, &\text{if~$U_i$~first~decodes~$x_2$}.
  \end{array}
  \right.
\end{align}
Specifically, if $\omega_1(t)=1$, $U_1$ decodes $x_1$ from its received signal by treating $x_2$ as interference, of which the signal-to-interference-plus-noise-ratio (SINR) can be expressed as
\begin{align}\label{Gamma_U1x1}
  \Gamma_{U_1\rightarrow x_1}^{\pi(t),1}=
  \frac{p_1(t)|h_1(t)|^2}{p_2(t)|h_1(t)|^2+\sigma^2},
\end{align}
where $\Gamma_{U_i\rightarrow x_j}^{\pi(t),\omega_i(t)}$ with $i,j\in\{1,2\}$ is used to denote the SINR/signal-to-noise-ratio (SNR) of $U_i$ to decode $x_j$ under decisions $\pi(t)$ and $\omega_i(t)$.
Moreover, if $\omega_1(t)=2$, $U_1$ first decodes $x_2$ by treating $x_1$ as interference, of which the SINR is expressed as
\begin{align}\label{Gamma_U1x2}
  \Gamma_{U_1\rightarrow x_2}^{\pi(t),2} =
  \frac{p_2(t)|h_1(t)|^2}{p_1(t)|h_1(t)|^2+\sigma^2}.
\end{align}
Then, by performing SIC with the obtained $x_2$, $U_1$ decodes its desired message $x_1$ with SNR
\begin{align}\label{gamma_U1x1}
  \Gamma_{U_1\rightarrow x_1}^{\pi(t),2} =
  \frac{p_1(t)|h_1(t)|^2}{\sigma^2}.
\end{align}
By denoting $\gamma_1\triangleq 2^{R_1}-1$ and $\gamma_2\triangleq 2^{R_2}-1$ as the SINR thresholds for decoding $x_1$ and $x_2$, respectively, the successful decoding event of $U_1$ can be expressed as
\begin{align}\label{Event_U1_Suc}
  \mathbb S^{\pi(t), \omega_1(t)}_1=
  \left\{
  \begin{array}{ll}
  \{\Gamma_{U_1\rightarrow x_1}^{\pi(t),1} \geq \gamma_1\}, &\text{if}~ \omega_1(t)=1,\\
  \{\Gamma_{U_1\rightarrow x_2}^{\pi(t),2} \geq \gamma_2, \Gamma_{U_1\rightarrow x_1}^{\pi(t),2} \geq \gamma_1\}, &\text{if}~ \omega_1(t)=2.
  \end{array}
  \right.
\end{align}
Similarly, by selecting $\omega_2(t)$ from $\{1,2\}$, the successful decoding event of $U_2$ is expressed as
\begin{align}\label{Event_U2_Suc}
  \mathbb S^{\pi(t), \omega_2(t)}_2=
  \left\{
  \begin{array}{ll}
  \{\Gamma_{U_2 \rightarrow x_2}^{\pi(t),2} \geq \gamma_2\}, &\text{if}~ \omega_2(t)=2,\\
  \{\Gamma_{U_2 \rightarrow x_1}^{\pi(t),1} \geq \gamma_1, \Gamma_{U_2 \rightarrow x_2}^{\pi(t),1} \geq \gamma_2\}, &\text{if}~ \omega_2(t)=1,
  \end{array}
  \right.
\end{align}
where $\Gamma_{U_2 \rightarrow x_2}^{\pi(t),2}$, $\Gamma_{U_2 \rightarrow x_1}^{\pi(t),1}$, and $\Gamma_{U_2 \rightarrow x_2}^{\pi(t),1}$ are respectively given by
\begin{align}\label{Gamma_U2}
  \Gamma_{U_2 \rightarrow x_2}^{\pi(t),2} =
  \frac{p_2(t)|h_2(t)|^2}{p_1(t)|h_2(t)|^2+\sigma^2},
  \Gamma_{U_2 \rightarrow x_1}^{\pi(t),1} =
  \frac{p_1(t)|h_2(t)|^2}{p_2(t)|h_2(t)|^2+\sigma^2},
  ~\text{and}~
  \Gamma_{U_2 \rightarrow x_2}^{\pi(t),1} =
  \frac{p_2(t)|h_2(t)|^2}{\sigma^2}.
\end{align}

Recall that the main aim of NOMA is to simultaneously serve multiple users on the same resource block, and thus the system outage probability is adopted as the performance metric in this paper. Using (\ref{Event_U1_Suc}) and (\ref{Event_U2_Suc}), the system outage probability is formulated as
\begin{align}\label{Form_Psys}
   \mathcal P_{\text{sys}}
   =1-\mathcal P\{ \mathbb S^{\pi(t), \omega_1(t)}_1 , \mathbb S^{\pi(t), \omega_2(t)}_2\}.
\end{align}
Furthermore, from the perspective of each user, the user outage probability of $U_i$ is written as
\begin{align}\label{Form_Pi}
   \mathcal P_{i}
   =1-{\mathcal P}\{ \mathbb S^{\pi(t), \omega_i(t)}_i\}, \text{for}~i=1,2.
\end{align}
For clarity, in the rest of this paper we use $[\pi(t), \omega_1(t), \omega_2(t)]$ to denote the joint decision on PA-DOS. As can be seen from (\ref{Gamma_U1x1})--(\ref{Form_Pi}), $[\pi(t), \omega_1(t), \omega_2(t)]$ significantly affects the system outage performance and users' outage performance.

\section{The Impact of Target Rates and the Designed PA-DOS Strategies}\label{Section_strategies}
In this section, we first investigate the impact of target rates on PA-DOS. It will be shown that, the available PA-DOS decisions are distinct under various target rates. Applying this result, two rate-adaptive criteria for PA-DOS are further developed, which reveal the universal properties of F-NOMA under arbitrary PA-DOS strategies. On the other hand, by employing the properties of F-NOMA, we propose two outage-optimal strategies with considering different user priorities. In the rest of this paper, the index $t$ of transmission blocks will be omitted for simplicity, if it does not cause confusion.

\subsection{Available PA-DOS Decisions}
Recall that, in NOMA networks, the message which is first decoded by one user must be detected with interference. To be specific, for either $U_i$ with $i\in\{1,2\}$, if $\omega_i=1$, $x_1$ is decoded with SINR $\Gamma_{U_i\rightarrow x_1}^{\pi,1}$, while if $\omega_i=2$, $x_2$ is decoded with SINR $\Gamma_{U_i\rightarrow x_2}^{\pi,2}$. As can be seen from (\ref{Gamma_U1x1}), (\ref{Gamma_U1x2}), and (\ref{Gamma_U2}), for any $i\in\{1,2\}$, $\Gamma_{U_i\rightarrow x_1}^{\pi,1}$ and $\Gamma_{U_i\rightarrow x_2}^{\pi,2}$ are monotone increasing functions of $p|h_i|^2$, and their limits with respect to $p|h_i|^2\rightarrow \infty$ are
\begin{align}\label{lim_Gamma1}
  \lim_{p|h_i|^2\rightarrow \infty} \Gamma_{U_i\rightarrow x_1}^{\pi,1}
  =\left\{
   \begin{array}{ll}
   \frac{\alpha}{1-\alpha}, \text{if}~\pi=1,\\
   \frac{1-\alpha}{\alpha}, \text{if}~\pi=2,
   \end{array}
   \right.
\end{align}\vspace{-3mm}
and
\begin{align}\label{lim_Gamma2}
  \lim_{p|h_i|^2\rightarrow \infty} \Gamma_{U_i\rightarrow x_2}^{\pi,2}
  =\left\{
   \begin{array}{ll}
   \frac{1-\alpha}{\alpha}, \text{if}~\pi=1,\\
   \frac{\alpha}{1-\alpha}, \text{if}~\pi=2.
   \end{array}
   \right.
\end{align}
Further, if $\lim_{p|h_i|^2\rightarrow \infty} \Gamma_{U_i\rightarrow x_1}^{\pi,1}\leq \gamma_1$, $\Gamma_{U_i\rightarrow x_1}^{\pi,1}<\gamma_1$ always holds. According to (\ref{Event_U1_Suc}) and (\ref{Event_U2_Suc}), selecting $\omega_i=1$ in this case leads to $\mathbb S^{\pi, \omega_i}_i=\mathbb S^{\pi, 1}_i=\mathbb I$ for any $i\in\{1,2\}$, where $\mathbb I$ denotes the impossible event, and $\mathbb S^{\pi, \omega_i}_i=\mathbb I$ indicates inevitable system outage and user outage of $U_i$. On the contrary, if $\lim_{p|h_i|^2\rightarrow \infty} \Gamma_{U_i\rightarrow x_1}^{\pi,1}> \gamma_1$, $\mathbb S^{\pi, 1}_i$ happens as long as $p|h_i|^2$ is large enough, i.e., $\mathbb S^{\pi, 1}_i\neq \mathbb I$. Thus, in this case we say $\omega_i=1$ is an available decoding order. Similarly, if $\lim_{p|h_i|^2\rightarrow \infty} \Gamma_{U_i\rightarrow x_2}^{\pi,2}\leq\gamma_2$, $\mathbb S^{\pi, 2}_i=\mathbb I$ for any $i=1,2$. Namely, the condition for decoding order $\omega_i=2$ to be available is $\lim_{p|h_i|^2\rightarrow \infty} \Gamma_{U_i\rightarrow x_2}^{\pi,2}>\gamma_2$.

As shown by (\ref{lim_Gamma1}) and (\ref{lim_Gamma2}), whether $\lim_{p|h_i|^2\rightarrow \infty} \Gamma_{U_i\rightarrow x_1}^{\pi,1}> \gamma_1$ and $\lim_{p|h_i|^2\rightarrow \infty} \Gamma_{U_i\rightarrow x_2}^{\pi,2}>\gamma_2$ hold is unrelated to user index $i$, but is determined by the PA decision $\pi$ and the system parameters $\gamma_1$, $\gamma_2$, and $\alpha$. Consequently, when $\pi$,  $\gamma_1$, $\gamma_2$, and $\alpha$ are given, the available decoding orders for $U_1$ and $U_2$ are always the same. For clarity, we define $\mathbf O(\pi)$ as the set which collects all the available decoding orders for $U_1$ and $U_2$, i.e., $\mathbf O(\pi)\triangleq \{j: j\in\{1,2\}, \lim_{p|h_i|^2\rightarrow \infty} \Gamma_{U_i\rightarrow x_j}^{\pi,j}> \gamma_j\}$. Employing (\ref{lim_Gamma1}) and (\ref{lim_Gamma2}), the detailed expression of $\mathbf O(\pi)$ can be obtained as
\begin{align}\label{Set_O}
    \left\{
    \begin{array}{lll}
    \mathbf O(1)=\emptyset, &\mathbf O(2)=\emptyset, &\text{if}~ (\gamma_1,\gamma_2)\in \mathbf R_0,\\
    \mathbf O(1)=\{1,2\}, &\mathbf O(2)=\{1,2\}, &\text{if}~ (\gamma_1,\gamma_2)\in \mathbf R_1,\\
    \mathbf O(1)=\{1,2\}, &\mathbf O(2)=\{2\}, &\text{if}~ (\gamma_1,\gamma_2)\in \mathbf R_2,\\
    \mathbf O(1)=\{2\}, &\mathbf O(2)=\{2\}, &\text{if}~ (\gamma_1,\gamma_2)\in \mathbf R_3,\\
    \mathbf O(1)=\{1\}, &\mathbf O(2)=\{2\}, &\text{if}~ (\gamma_1,\gamma_2)\in \mathbf R_4,\\
    \mathbf O(1)=\emptyset, &\mathbf O(2)=\{2\}, &\text{if}~ (\gamma_1,\gamma_2)\in \mathbf R_5,
    \end{array}
    \right.
\end{align}
\begin{figure}
  \centering
  \includegraphics[width=2.7in]{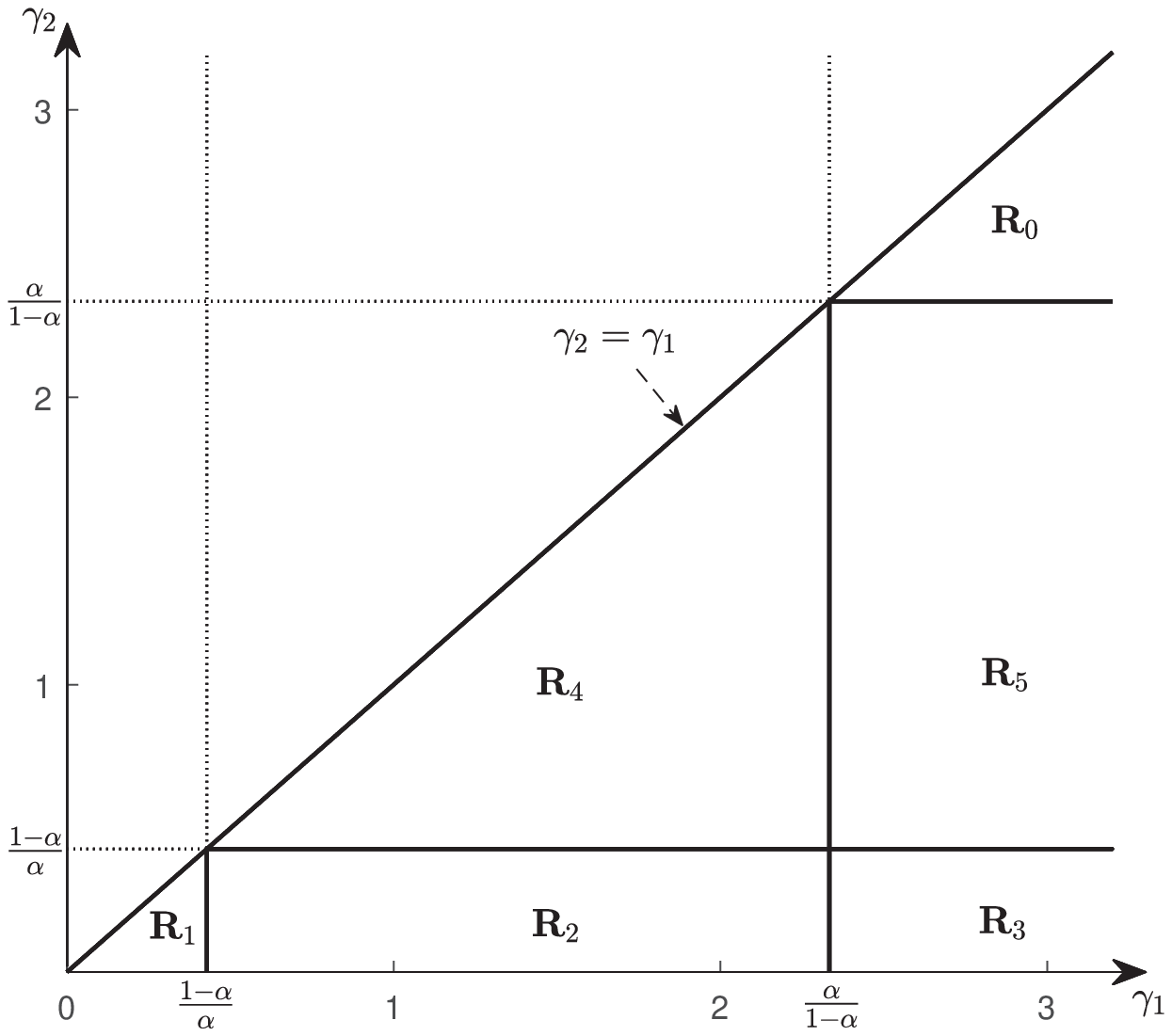}\\
  \caption{An illustration of $\mathbf R_k$, $k=0,1,...,5$, with $\alpha=0.7$.}\label{fig_Ava_regions}\vspace{-1cm}
\end{figure}
where $\mathbf R_k$, $k=0,1,...,5$, are a set of subregions defined as
\begin{align}\label{Region_Rk}
  \left\{
  \begin{array}{l}
    \mathbf R_0 \triangleq \{(\gamma_1, \gamma_2):\gamma_1\geq\gamma_2\geq\tfrac{\alpha}{1-\alpha}>0 \}, \\
    \mathbf R_1 \triangleq \{(\gamma_1, \gamma_2):\tfrac{\alpha}{1-\alpha}\geq\tfrac{1-\alpha}{\alpha}>\gamma_1\geq\gamma_2>0 \}, \\
    \mathbf R_2 \triangleq \{(\gamma_1, \gamma_2):\tfrac{\alpha}{1-\alpha}>\gamma_1\geq\tfrac{1-\alpha}{\alpha}>\gamma_2>0 \}, \\
    \mathbf R_3 \triangleq \{(\gamma_1, \gamma_2):\gamma_1\geq\tfrac{\alpha}{1-\alpha}\geq\tfrac{1-\alpha}{\alpha}>\gamma_2>0 \}, \\
    \mathbf R_4 \triangleq \{(\gamma_1, \gamma_2):\tfrac{\alpha}{1-\alpha}>\gamma_1\geq\gamma_2\geq \tfrac{1-\alpha}{\alpha}>0 \}, \\
    \mathbf R_5 \triangleq \{(\gamma_1, \gamma_2):\gamma_1\geq\tfrac{\alpha}{1-\alpha}>\gamma_2\geq\tfrac{1-\alpha}{\alpha}>0 \},
  \end{array}
  \right.
\end{align}
which divide the region of $(\gamma_1,\gamma_2)$ bounded by $\gamma_1\geq\gamma_2>0$, as illustrated by Fig. \ref{fig_Ava_regions}.

Moreover, we define a set $\mathbf P$ to collect all the PA results which guarantee $\mathbf O(\pi)\neq\emptyset$, i.e.,  $\mathbf P\triangleq \{\pi: \pi\in\{1,2\}, \mathbf O(\pi)\neq\emptyset\}$. Since $\mathbf O(\pi)=\emptyset$ leads to $\mathbb S^{\pi, \omega_i}_i=\mathbb I$ for any $i\in\{1,2\}$ and $\omega_i\in\{1,2\}$, we call $\mathbf P$ the set of available PA decisions. Using (\ref{Set_O}), the detailed expression of $\mathbf P$ can be obtained as
\begin{align}\label{Set_P}
  \mathbf P=
  \left\{
  \begin{array}{ll}
  \emptyset, &\text{if}~ (\gamma_1,\gamma_2)\in \mathbf R_0, \\
  \{\pi_1,\pi_2\}, &\text{if}~ (\gamma_1,\gamma_2)\in \cup_{k=1,...,4}\mathbf R_k, \\
  \{\pi_2\},  &\text{if}~ (\gamma_1,\gamma_2)\in \mathbf R_5.
  \end{array}
  \right.
\end{align}


Further, by substituting (\ref{Gamma_U1x1})--(\ref{gamma_U1x1}) in (\ref{Event_U1_Suc}) and substituting (\ref{Gamma_U2}) in (\ref{Event_U2_Suc}) while employing the above defined $\mathbf P$ and $\mathbf O(\pi)$, event $\mathbb S^{\pi, \omega_i}_i$ can be further expressed as, for $i=1,2$,
\begin{align}\label{Event_Si}
  \mathbb S^{\pi, \omega_i}_i=
  \left\{
  \begin{array}{ll}
  \{\rho|h_i|^2 \geq \phi_i^{\pi,\omega_i} \}, &\text{if}~\pi\in\mathbf P, \omega_i\in \mathbf O(P),\\
  \mathbb I, &\text{otherwise},
  \end{array}
  \right.
\end{align}
where $\rho\triangleq p/\sigma^2$ denotes the transmit SNR, and $\phi_i^{\pi,\omega_i}$, $i=1,2$, are defined as
\begin{align}\label{phi_1}
  \phi_1^{\pi,\omega_1}
  &\triangleq\left\{
    \begin{array}{ll}
    \frac{\gamma_1}{\alpha-(1-\alpha)\gamma_1}, &\text{if}~\pi=1, \omega_1=1,\\
    \max \{\frac{\gamma_2}{1-\alpha-\alpha\gamma_2},\frac{\gamma_1}{\alpha}\}, &\text{if}~\pi=1, \omega_1=2,\\
    \frac{\gamma_1}{1-\alpha-\alpha\gamma_1}, &\text{if}~\pi=2, \omega_1=1,\\
    \max \{\frac{\gamma_2}{\alpha-(1-\alpha)\gamma_2},\frac{\gamma_1}{1-\alpha}\}, &\text{if}~\pi=2, \omega_1=2,
    \end{array}
    \right.
\end{align}\vspace{-3mm}
and
\begin{align}\label{phi_2}
  \phi_2^{\pi,\omega_2}
  &\triangleq\left\{
    \begin{array}{ll}
    \max\{\frac{\gamma_1}{\alpha-(1-\alpha)\gamma_1},\frac{\gamma_2}{1-\alpha}\}, &\text{if}~\pi=1, \omega_2=1,\\
    \frac{\gamma_2}{1-\alpha-\alpha\gamma_2}, &\text{if}~\pi=1, \omega_2=2,\\
    \max \{\frac{\gamma_1}{1-\alpha-\alpha\gamma_1},\frac{\gamma_2}{\alpha}\}, &\text{if}~\pi=2, \omega_2=1,\\
    \frac{\gamma_2}{\alpha-(1-\alpha)\gamma_2}, &\text{if}~\pi=2, \omega_2=2,
    \end{array}
    \right.
\end{align}
which satisfy $\phi_i^{\pi,\omega_i}>0$ if $\pi\in\mathbf P$ and $\omega_i\in \mathbf O(\pi)$.
\vspace{-3mm}
\subsection{The Developed Rate-Adaptive Criteria for PA-DOS}\vspace{-3mm}
In this subsection, we present two rate-adaptive criteria for PA-DOS, which are regarding the system outage performance and the user outage performance, respectively.

\subsubsection{The Rate-Adaptive Criterion for PA-DOS on System Outage Performance}
By applying (\ref{Event_Si}) in (\ref{Form_Psys}) with some manipulations, the system outage probability can be rewritten as\vspace{-3mm}
\begin{align}\label{Psys_re}
  \mathcal P_{\text{sys}}
  ={\mathcal P}\{ \Phi^{\pi,\omega_1,\omega_2} < 1\},
\end{align}
where $\Phi^{\pi,\omega_1,\omega_2}$ is defined as
\begin{align}\label{Def_Phi}
  \Phi^{\pi,\omega_1,\omega_2} \triangleq
  \left\{
  \begin{array}{ll}
  \min \{ \rho|h_1|^2 / \phi_1^{\pi,\omega_1} , \rho|h_2|^2 / \phi_2^{\pi,\omega_2} \}, &\text{if}~\pi\in\mathbf P, \omega_1,\omega_2\in \mathbf O(\pi),\\
  0, &\text{otherwise}.
  \end{array}
  \right.
\end{align}
As shown by (\ref{Psys_re}), the system outage probability is expressed as a probability with regard to the value of $\Phi^{\pi,\omega_1,\omega_2}$, which is determined by random variables $|h_1|^2$ and $|h_2|^2$ as well as the PA-DOS decision $[\pi,\omega_1,\omega_2]$. Based on such a connection between $\mathcal P_{\text{sys}}$ and $[\pi,\omega_1,\omega_2]$, we have the following lemma, where $\mathcal A$ is used to denote an arbitrary PA-DOS strategy, $[\pi^{\mathcal A},\omega_1^{\mathcal A},\omega_2^{\mathcal A}]$ denotes the PA-DOS decision made by $\mathcal A$, $\mathcal P_{\text{sys}}^{\mathcal A}$ denotes the system outage probability achieved by $\mathcal A$, and ${{\mathcal P}}^{\text{min}}_{\text{sys}}$ represents the minimum system outage probability.

\begin{lemma}\label{lemma_NScondition}
  The necessary and sufficient condition for one strategy $\mathcal A$ to achieve the minimum system outage probability (${\mathcal P}^{\mathcal A}_{\text{sys}}={\mathcal P}^{\text{min}}_{\text{sys}}$) is $\mathcal P\{\Phi^{\pi^{\mathcal A},\omega_1^{\mathcal A},\omega_2^{\mathcal A}}<1,\max_{\pi,\omega_1,\omega_2\in\{1,2\}}\Phi^{\pi,\omega_1,\omega_2}\geq 1\}=0$.
\end{lemma}
\begin{IEEEproof}
  Please refer to Appendix \ref{Proof_lemma_NScondition}.
\end{IEEEproof}

Lemma \ref{lemma_NScondition} establishes a connection between the maximum $\Phi^{\pi,\omega_1,\omega_2}$ and the outage-optimal strategy. According to (\ref{Def_Phi}), the maximum value of $\Phi^{\pi,\omega_1,\omega_2}$ can be achieved as long as the smaller term between $\rho|h_1|^2 / \phi_1^{\pi,\omega_1}$ and $\rho|h_2|^2 / \phi_2^{\pi,\omega_2}$ is maximized. Thus, it is possible to exist more than one $[\pi, \omega_1, \omega_2]$ that can achieve the maximum $\Phi^{\pi,\omega_1,\omega_2}$ at each transmission block. Here, we present a general result (the result is always true) of $\max_{\pi,\omega_1,\omega_2\in\{1,2\}}\Phi^{\pi,\omega_1,\omega_2}$ as follows:

\begin{lemma}\label{lemma_maxPhi}
   A general result of $\max_{\pi,\omega_1,\omega_2\in\{1,2\}} \Phi^{\pi,\omega_1,\omega_2}$ can be given as
  \begin{align}\label{max_Phi}
    \max_{\pi,\omega_1,\omega_2\in\{1,2\}} \Phi^{\pi,\omega_1,\omega_2}
    =\left\{
    \begin{array}{ll}
    0, &\text{if}~(\gamma_1,\gamma_2)\in \mathbf R_0, \\
    \Phi^{1,1,1}, &\text{if}~(\gamma_1,\gamma_2)\in \mathbf S_2, {|h_1|^2}/{|h_2|^2}<{\phi_1^{2,2}}/{\phi_2^{1,1}}, \\
    \Phi^{1,2,2}, &\text{if}~(\gamma_1,\gamma_2)\in \mathbf S_3, {|h_1|^2}/{|h_2|^2}<{\phi_1^{2,2}}/{\phi_2^{1,2}}, \\
    \Phi^{2,2,2}, &\text{otherwise},
    \end{array}
    \right.
  \end{align}
  where $\mathbf S_1$, $\mathbf S_2$, and $\mathbf S_3$ are defined as
  \begin{align}\label{Def_R_HUF}
  \left\{
  \begin{array}{ll}
  \mathbf S_1 &\triangleq  \big\{ (\gamma_1, \gamma_2): \gamma_1 \geq \gamma_2 >0, \gamma_1\geq \tfrac{\alpha}{1-\alpha}-1, \tfrac{\alpha}{1-\alpha}> \gamma_2\geq \tfrac{(1-\alpha)\gamma_1}{ 1-\alpha+\alpha\gamma_1}  \big\}, \\
  \mathbf S_2 &\triangleq  \big\{ (\gamma_1, \gamma_2): \gamma_1 \geq \gamma_2 >0, \gamma_1< \tfrac{\alpha}{1-\alpha}-1, \gamma_2\geq \tfrac{(1-\alpha)\gamma_1}{\alpha+(2\alpha-1)\gamma_1} \big\},\\
  \mathbf S_3 &\triangleq  \big\{ (\gamma_1, \gamma_2): \gamma_1 \geq \gamma_2 >0, \gamma_2< \min\big\{ \tfrac{(1-\alpha)\gamma_1}{\alpha+(2\alpha-1)\gamma_1} , \tfrac{(1-\alpha)\gamma_1} {1-\alpha+\alpha\gamma_1} \big\}\big\},
  \end{array}
  \right.
  \end{align}
  which satisfy $\cup_{k=1,2,3} \mathbf S_k=\mathbf R^{\text{ava}}\triangleq \cup_{k=1,2,...,5} \mathbf R_k$, as illustrated in Fig. \ref{fig_HUF_regions}.
\end{lemma}
\begin{IEEEproof}
  Please refer to Appendix \ref{Proof_best_decision}.
\end{IEEEproof}

Based on Lemma \ref{lemma_NScondition} and Lemma \ref{lemma_maxPhi}, we propose the rate-adaptive criterion of PA-DOS on system outage performance, which is given by the following proposition.
\begin{proposition}\label{Prop_cond_sys}
  First, when $(\gamma_1, \gamma_2)\in\mathbf R_0$, we have $\mathcal P_{\text{sys}}^{\text{min}}=1$, namely the system outage inevitably occurs for any PA-DOS strategy; When $(\gamma_1, \gamma_2)\in\cup_{k=1,2,3}\mathbf S_k$, we have $\mathcal P_{\text{sys}}^{\text{min}}< 1$. Second, when $(\gamma_1, \gamma_2)\in\mathbf S_1$, there exists a channel-unrelated strategy $\mathcal A^*$ as $[\pi^{\mathcal A^*},\omega_1^{\mathcal A^*},\omega_2^{\mathcal A^*}]=[2,2,2]$, which achieves $\mathcal P_{\text{sys}}^{\mathcal A^*}=\mathcal P_{\text{sys}}^{\text{min}}$. Third, when $(\gamma_1, \gamma_2)\in\mathbf S_2\cup\mathbf S_3$, $\mathcal P_{\text{sys}}^{\text{min}}$ cannot be achieved by channel-unrelated strategies.
\end{proposition}
\begin{IEEEproof}
  Please refer to Appendix \ref{Proof_Prop_cond_sys}.
\end{IEEEproof}

\subsubsection{The Rate-Adaptive Criterion for PA-DOS on User Outage Performance}
For clarity, in the following we use $\mathcal P_i^{\mathcal A}$ to denote $U_i$'s outage probability achieved by strategy $\mathcal A$, and use ${{\mathcal P}}^{\text{min}}_i$ to denote $U_i$'s minimum outage probability.

\begin{lemma}\label{Lemma_Pimin}\vspace{-3mm}
  The minimum outage probability of $U_i$, $i=1,2$, can be written as
  \begin{align} \label{P1min}
  {{\mathcal P}}^{\text{min}}_{1}&=
    \left\{
    \begin{array}{ll}
    1, &\text{if}~(\gamma_1,\gamma_2)\in\mathbf R_0,\\
    {\mathcal P}(\rho|h_1|^2 < \phi_1^{2,2}), &\text{if}~(\gamma_1,\gamma_2)\in\mathbf S_1, \\
    {\mathcal P}(\rho|h_1|^2 < \phi_1^{1,1}), &\text{if}~(\gamma_1,\gamma_2)\in\mathbf S_2, \\
    {\mathcal P}(\rho|h_1|^2 < \phi_1^{1,2}), &\text{if}~(\gamma_1,\gamma_2)\in\mathbf S_3,
    \end{array}
    \right.
  \end{align}\vspace{-3mm}
  and
  \begin{align}\label{P2min}
  {{\mathcal P}}^{\text{min}}_{2}&=
  \left\{
  \begin{array}{ll}
  1, &\text{if}~(\gamma_1,\gamma_2)\in\mathbf R_0,\\
  {\mathcal P}(\rho|h_2|^2 < \phi_2^{2,2}), &\text{if}~(\gamma_1,\gamma_2)\in\mathbf R^{\text{ava}}.
  \end{array}
  \right.
  \end{align}
\end{lemma}
\begin{IEEEproof}
  As revealed by (\ref{Form_Pi}), (\ref{Set_O}), (\ref{Set_P}), and (\ref{Event_Si}), when $(\gamma_1,\gamma_2)\in\mathbf R_0$, $\mathcal P_i=1$ for any $\pi,\omega_i\in\{1,2\}$, i.e., ${{\mathcal P}}^{\text{min}}_{i}=1$. Correspondingly, when $(\gamma_1,\gamma_2)\in\mathbf R^{\text{ava}}$, $\pi$ and $\omega_i$ should be selected from $\mathbf P$ and $\mathbf O(\pi)$ to achieve ${{\mathcal P}}^{\text{min}}_{i}$, which leads to ${{\mathcal P}}_{i}={{\mathcal P}}\{\rho|h_i|^2 < \phi_i^{\pi,\omega_i} \}$. Further, combining (\ref{O_iprime_P1}), (\ref{O_iprime_P2}), (\ref{Ineq_phii11_phii22}), and (\ref{Ineq_phii12_phii22}) it can be known that, $\min_{\pi\in\mathbf P,\omega_2\in\mathbf O(\pi)}\phi_2^{\pi,\omega_2}=\phi_2^{2,2}$ holds for $(\gamma_1,\gamma_2)\in\mathbf R^{\text{ava}}$, while $\min_{\pi\in\mathbf P,\omega_1\in\mathbf O(\pi)}\phi_1^{\pi,\omega_1}$ equals $\phi_1^{2,2}$, $\phi_1^{1,1}$, and $\phi_1^{1,2}$, when $(\gamma_1,\gamma_2)\in\mathbf S_1,\mathbf S_2,\mathbf S_3$, respectively. Combining these facts with ${\mathcal P}(\rho|h_i|^2 < \min_{\pi\in\mathbf P,\omega_i\in\mathbf O(\pi)}\phi_i^{\pi,\omega_i})\leq {\mathcal P}(\rho|h_i|^2 < \phi_i^{\pi,\omega_i})$ for any  $i\in\{1,2\}$, $\pi\in\mathbf P$, and $\omega_i\in\mathbf O(\pi)$, (\ref{P1min}) and (\ref{P2min}) are obtained.
\end{IEEEproof}

Based on Lemma \ref{Lemma_Pimin}, we have the following rate-adaptive criterion of PA-DOS on user outage performance.
\begin{proposition}\label{Prop_cond_user}
  First, when $(\gamma_1, \gamma_2)\in\mathbf R_0$, we have $\mathcal P_i^{\text{min}}=1$ for $i=1,2$, namely the user outage inevitably occurs for any PA-DOS strategy. Second, when $(\gamma_1, \gamma_2)\in\mathbf S_1$, the channel-unrelated strategy $\mathcal A^*$ with $[\pi^{\mathcal A^*},\omega_1^{\mathcal A^*},\omega_2^{\mathcal A^*}]=[2,2,2]$ simultaneously achieves $\mathcal P_1^{\mathcal A^*}=\mathcal P_1^{\text{min}}$ and $\mathcal P_2^{\mathcal A^*}=\mathcal P_2^{\text{min}}$. Third, when $(\gamma_1, \gamma_2)\in\mathbf S_2\cup\mathbf S_3$, $\mathcal P_1^{\text{min}}$ and $\mathcal P_2^{\text{min}}$ cannot be simultaneously achieved by any PA-DOS strategy.
\end{proposition}
\begin{IEEEproof}
  Please refer to Appendix \ref{Proof_Prop_cond_user}.
\end{IEEEproof}

From Proposition \ref{Prop_cond_sys} and Proposition \ref{Prop_cond_user}, the following important universal properties of F-NOMA can be summarized:

First, when $(\gamma_1, \gamma_2)\in \mathbf R_0$, the system outage and the user outage inevitably occur. Thus, the operation condition for F-NOMA is $(\gamma_1, \gamma_2)\in \mathbf R^\text{ava}$, i.e., $R_2<\log_2(1+\alpha/(1-\alpha))$. It means that, F-NOMA can be implemented to arbitrary two users as long as either of them requests a target rate smaller than $\log_2(1+\alpha/(1-\alpha))$ \footnote{It is worth noting that, similar operation conditions are also defined in existing works with certain PA-DOS strategies \cite{Ding_random_users,YLiu_CoNOMA}, which stipulate the target rate of one specific user to be lower than $\log_2(1+\alpha/(1-\alpha))$, e.g., the weak user is stipulated in the CSD strategy \cite{Ding_random_users} and the far user is stipulated in the DD strategy \cite{YLiu_CoNOMA}. As revealed by the proposed operation condition, the existing conditions are over restricted. As the result, under some system parameters, the outage error floors caused by the existing strategies can actually be avoided by properly conducting PA-DOS, as will be shown in Sections \ref{subsection_Psys} and \ref{Section_simulation}.};

Second, when the target rates satisfy $(\gamma_1, \gamma_2)\in\mathbf S_1$, allocating the higher power to the low-rate user and decoding its message first is not only the outage-optimal PA-DOS decision for both the system and the users, but is also signaling-efficient due to its independence from the CSI;

At last, when $(\gamma_1, \gamma_2)\in\mathbf S_2\cup\mathbf S_3$, CSIT is crucial for PA-DOS to achieve the minimum system outage probability. Meanwhile, the minimum user outage probabilities cannot be simultaneously achieved for the two users. This can be intuitively explained as follows. As indicated by Lemma \ref{Lemma_Pimin}, when $(\gamma_1, \gamma_2)\in\mathbf S_2\cup\mathbf S_3$, the two users both desire the higher power for their best outage performance, whereas the higher power can only be allocated to one user in one transmission block. We refer to this phenomenon as \emph{the power contradiction} of F-NOMA for case $(\gamma_1, \gamma_2)\in\mathbf S_2\cup\mathbf S_3$.

\subsection{The Proposed HUF and LUF Strategies}
Since when $(\gamma_1, \gamma_2)\in \mathbf R_0$ the system outage and the user outage occur for sure, in this subsection we focus on the design of PA-DOS strategies for case $(\gamma_1, \gamma_2)\in \mathbf R^\text{ava}$. Following Proposition \ref{Prop_cond_sys} and Proposition \ref{Prop_cond_user} while considering different priorities of $U_1$ and $U_2$, we propose two strategies termed as HUF and LUF.

For the case that the high-rate user $U_1$ has the higher priority, the HUF strategy is developed as: 1) When $(\gamma_1, \gamma_2)\in \mathbf S_1$, $[\pi^{\mathcal H},\omega_1^{\mathcal H},\omega_2^{\mathcal H}]=[2,2,2]$; 2) When $(\gamma_1, \gamma_2)\in \mathbf S_2$, $[\pi^{\mathcal H},\omega_1^{\mathcal H},\omega_2^{\mathcal H}]= [2,2,2]$ if $|h_1|^2 \geq \frac{{\phi_1^{2,2}}}{\rho}$, and $[\pi^{\mathcal H},\omega_1^{\mathcal H},\omega_2^{\mathcal H}]=[1,1,1]$ otherwise; 3) When $(\gamma_1, \gamma_2)\in \mathbf S_3$, $[\pi^{\mathcal H},\omega_1^{\mathcal H},\omega_2^{\mathcal H}]=[2,2,2]$ if $|h_1|^2 \geq \frac{{\phi_1^{2,2}}}{\rho}$, and $[\pi^{\mathcal H},\omega_1^{\mathcal H},\omega_2^{\mathcal H}]=[1,2,2]$ otherwise. By comparing the HUF strategy with (\ref{Set_O}) and (\ref{Set_P}) it can be known that, $[\pi^{\mathcal H},\omega_1^{\mathcal H},\omega_2^{\mathcal H}]$ must satisfy $\pi^{\mathcal H}\in\mathbf P$ and $\omega_1^{\mathcal H},\omega_2^{\mathcal H}\in\mathbf O(\pi^{\mathcal H})$.

\begin{corollary}\label{Coro_HUF}
  When $(\gamma_1, \gamma_2)\in \mathbf R^{\text{ava}}$, the HUF strategy achieves the minimum system outage probability, i.e., $\mathcal P_{\text{sys}}^{\mathcal H}=\mathcal P_{\text{sys}}^{\text{min}}$. On the other hand, when $(\gamma_1, \gamma_2)\in \mathbf S_1$, the HUF strategy achieves the minimum user outage probabilities for both $U_1$ and $U_2$, i.e., $\mathcal P_i^{\mathcal H}=\mathcal P_i^{\text{min}}$ for $i=1,2$; When $(\gamma_1, \gamma_2)\in \mathbf S_2\cup\mathbf S_3$, it achieves the minimum user outage probability for $U_1$, i.e., $\mathcal P_1^{\mathcal H}=\mathcal P_1^{\text{min}}$.
\end{corollary}
\begin{IEEEproof}
  Please refer to Appendix \ref{Proof_Coro_HUF}.
\end{IEEEproof}

For the case that the low-rate user $U_2$ has the higher priority, the LUF strategy is developed as: 1) When $(\gamma_1, \gamma_2)\in \mathbf S_1$, $[\pi^{\mathcal L},\omega_1^{\mathcal L},\omega_2^{\mathcal L}]=[2,2,2]$; 2) When $(\gamma_1, \gamma_2)\in \mathbf S_2$, $[\pi^{\mathcal L},\omega_1^{\mathcal L},\omega_2^{\mathcal L}]=[2,2,2]$ if $|h_2|^2 < {\phi_2^{1,1}}/{\rho}$, and $[\pi^{\mathcal L},\omega_1^{\mathcal L},\omega_2^{\mathcal L}]=[1,1,1]$ otherwise; 3) When $(\gamma_1, \gamma_2)\in \mathbf S_3$, $[\pi^{\mathcal L},\omega_1^{\mathcal L},\omega_2^{\mathcal L}]=[2,2,2]$ if $|h_2|^2 < {{\phi_2^{1,2}}}/{\rho}$, and $[\pi^{\mathcal L},\omega_1^{\mathcal L},\omega_2^{\mathcal L}]=[1,2,2]$ otherwise. Similar to the HUF strategy, $[\pi^{\mathcal L},\omega_1^{\mathcal L},\omega_2^{\mathcal L}]$ must satisfy $\pi^{\mathcal L}\in\mathbf P$ and $\omega_1^{\mathcal L},\omega_2^{\mathcal L}\in\mathbf O(\pi^{\mathcal L})$.
\begin{corollary}\label{Coro_LUF}
  When $(\gamma_1, \gamma_2)\in \mathbf R^{\text{ava}}$, the LUF strategy also achieves the minimum system outage probability, i.e., $\mathcal P_{\text{sys}}^{\mathcal L}=\mathcal P_{\text{sys}}^{\text{min}}$. On the other hand, when $(\gamma_1, \gamma_2)\in \mathbf S_1$, we have $\mathcal P_i^{\mathcal L}=\mathcal P_i^{\text{min}}$ for $i=1,2$; When $(\gamma_1, \gamma_2)\in \mathbf S_2\cup\mathbf S_3$, LUF achieves the minimum user outage probability for $U_2$, i.e., $\mathcal P_2^{\mathcal L}=\mathcal P_2^{\text{min}}$.
\end{corollary}
\begin{IEEEproof}
  Following the same rationale in Appendix \ref{Proof_Coro_HUF}, Corollary \ref{Coro_LUF} can be proved.
\end{IEEEproof}

As can be seen, when $(\gamma_1,\gamma_2)$ respectively locates in $\mathbf S_1$, $\mathbf S_2$, and $\mathbf S_3$, distinct PA-DOS decisions are made by the proposed strategies, which result in different complexity for the implementation of PA-DOS. Specifically, if $(\gamma_1,\gamma_2)\in \mathbf S_1$, the PA-DOS decision is channel-unrelated, and it only needs to be informed once before the entire data transmission process. On the contrary, if $(\gamma_1, \gamma_2)\in \mathbf S_2\cup\mathbf S_3$, the PA-DOS decision is channel-related. In this case, the HUF strategy can be conducted based on the one-bit feedback from $U_1$, which indicates whether $|h_1|^2\geq\phi_1^{2,2}/\rho$. Similarly, the LUF strategy can be implemented with one-bit feedback from $U_2$, which indicates whether $|h_2|^2 < {\phi_2^{1,1}}/{\rho}$ for case $(\gamma_1, \gamma_2)\in \mathbf S_2$, and whether $|h_2|^2 < {{\phi_2^{1,2}}}/{\rho}$ for case $(\gamma_1, \gamma_2)\in \mathbf S_3$. As the summery, at most one time of computation and one bit of signaling are required for PA-DOS at each transmission block.





\section{Performance Analysis}\label{Section_performance}
In this section, we first derive the exact and high-SNR asymptotic results of the system outage probabilities achieved by the proposed strategies. Meanwhile, to verify the effectiveness of the proposed strategies, the performance gain over the most widely adopted CSD strategy is theoretically evaluated. Further, we derive the user outage probabilities achieved by the proposed strategies. The results show that, the impact of the power contradiction on the user outage performance is negligible in the high-SNR regime when implementing the proposed strategies.
\subsection{System Outage Probabilities Achieved by the Proposed Strategies}\label{subsection_Psys}
With adopting Rayleigh fading, channel gains $|h_i(t)|^2$, $i=1,2$, are assumed to follow the exponential distribution with means equal to $\eta_i$, $i=1,2$. Here, $\eta_i$ is used to characterize the pathloss of the wireless channel, which is modeled as $\eta_i=(1+d_i/d_0)^{-\nu}$, with $d_i$ denoting the distance from the BS to $U_i$, $d_0$ denoting the reference distance, and $\nu$ denoting the pathloss exponent. Correspondingly, the probability density function (PDF) of $|h_i|^2$, $i=1,2$, can be expressed as $f_{|h_i|^2}(x)=\frac{1}{\eta_i}e^{-x/\eta_i}$ for $x\geq 0$.

Recall that, when $(\gamma_1,\gamma_2)\in\mathbf R_0$, system outage inevitably occurs, and when $(\gamma_1,\gamma_2)\in\mathbf R^{\text{ava}}$, the decision of the HUF strategy, i.e., $[\pi^{\mathcal H},\omega_1^{\mathcal H},\omega_2^{\mathcal H}]$, must satisfy $\pi^{\mathcal H}\in\mathbf P$ and $\omega_1^{\mathcal H},\omega_2^{\mathcal H}\in\mathbf O(\pi^{\mathcal H})$. Based on theses facts, using (\ref{Form_Psys}) and (\ref{Event_Si}) the system outage probability achieved by HUF can be written as
\begin{align}\label{Pout_HUF_sys_formu}
  {\mathcal P}^{\mathcal H}_{\text{sys}}
  =\left\{
  \begin{array}{ll}
  1, &\text{if}~(\gamma_1,\gamma_2)\in\mathbf R_0,\\
  1-\mathcal P\{ \rho|h_1|^2\geq\phi_1^{\pi^{\mathcal H}, \omega_1^{\mathcal H}} , \rho|h_2|^2\geq\phi_2^{\pi^{\mathcal H}, \omega_2^{\mathcal H}}\}, &\text{if}~ (\gamma_1,\gamma_2)\in\mathbf R^{\text{ava}}.
  \end{array}
  \right.
\end{align}
Then, employing the HUF strategy and the PDF of $|h_i|^2$ in (\ref{Pout_HUF_sys_formu}), ${\mathcal P}^{\mathcal H}_{\text{sys}}$ can be calculated as
\begin{align}\label{Pout_HUF_sys}
  {\mathcal P}^{\mathcal H}_{\text{sys}}
  =\left\{
  \begin{array}{ll}
  1, &\text{if}~ (\gamma_1, \gamma_2)\in\mathbf R_0, \\
  {\mathcal P}^{\mathcal H}_{\text{sys},1}, &\text{if}~ (\gamma_1, \gamma_2)\in \mathbf S_1, \\
  {\mathcal P}^{\mathcal H}_{\text{sys},2}, &\text{if}~ (\gamma_1, \gamma_2)\in \mathbf S_2, \\
  {\mathcal P}^{\mathcal H}_{\text{sys},3}, &\text{if}~ (\gamma_1, \gamma_2)\in \mathbf S_3,
  \end{array}
  \right.
\end{align}
where ${\mathcal P}^{\mathcal H}_{\text{sys},k}$, $k=1,2,3$, are defined and calculated as
\begin{align}\label{Pouts_HUF1}
  {\mathcal P}^{\mathcal H}_{\text{sys},1}
  \triangleq1- {{\mathcal P}}\{ \rho|h_1|^2 \geq \phi_1^{2,2}, \rho|h_2|^2 \geq \phi_2^{2,2} \}
  =1- e^{-\frac{1}{\rho}(\frac{1}{\eta_1}{\phi_1^{2,2}} + \frac{1}{\eta_2}{\phi_2^{2,2}} )},
\end{align}
\begin{align}\label{Pouts_HUF2}
  {\mathcal P}^{\mathcal H}_{\text{sys},2}
  \triangleq&1-\mathcal P\{\rho|h_1|^2\geq\phi_1^{2,2},\rho|h_2|^2\geq\phi_2^{2,2}\}-\mathcal P\{\phi_1^{2,2}>\rho|h_1|^2\geq\phi_1^{1,1},\rho|h_2|^2\geq\phi_2^{1,1}\}\notag\\
  =&1 -e^{-\frac{1}{\rho}(\frac{1}{\eta_1}{\phi_1^{2,2}}+\frac{1}{\eta_2}{\phi_2^{2,2}})}
      -e^{-\frac{1}{\rho}(\frac{1}{\eta_1}{\phi_1^{1,1}}+\frac{1}{\eta_2}{\phi_2^{1,1}})}
   +e^{-\frac{1}{\rho}(\frac{1}{\eta_1}{\phi_1^{2,2}}+\frac{1}{\eta_2}{\phi_2^{1,1}})},
\end{align}
and
\begin{align}\label{Pouts_HUF3}
  {{\mathcal P}}^{\mathcal H}_{\text{sys},3}
  \triangleq&1-\mathcal P\{\rho|h_1|^2\geq\phi_1^{2,2},\rho|h_2|^2\geq\phi_2^{2,2}\}-\mathcal P\{\phi_1^{2,2}>\rho|h_1|^2\geq\phi_1^{1,2},\rho|h_2|^2\geq\phi_2^{1,2}\}\notag\\
  =&1 - e^{- \frac{1}{\rho} (\frac{1}{\eta_1} {\phi_1^{2,2}} + \frac{1}{\eta_2} {\phi_2^{2,2}} )}
      - e^{- \frac{1}{\rho} (\frac{1}{\eta_1} {\phi_1^{1,2}} + \frac{1}{\eta_2} {\phi_2^{1,2}} )}
   + e^{-\frac{1}{\rho} (\frac{1}{\eta_1} {\phi_1^{2,2}} + \frac{1}{\eta_2} {\phi_2^{1,2}}) }.
\end{align}
In the above, the equality of (\ref{Pouts_HUF2}) is guaranteed by $\phi_1^{2,2}>\phi_1^{1,1}$ when $(\gamma_1, \gamma_2)\in \mathbf S_2$, as given by (\ref{Ineq_phii11_phii22}), and the equality of (\ref{Pouts_HUF3}) is guaranteed by $\phi_1^{2,2}>\phi_1^{1,2}$ when $(\gamma_1, \gamma_2)\in \mathbf S_3$, as given by (\ref{Ineq_phii12_phii22}).
Furthermore, by applying $1-e^{-x}\overset{x\rightarrow 0}{\simeq} x$ in (\ref{Pout_HUF_sys}), the high-SNR asymptotic result of ${{\mathcal P}}^{\mathcal H}_{\text{sys}}$ can be obtained as
\begin{align}\label{Pout_HUF_asym}
  {{\mathcal P}}^{\mathcal H}_{\text{sys}}
  \overset{\rho\rightarrow \infty}{\simeq}
  \tilde {{\mathcal P}}^{\mathcal H}_{\text{sys}}=
  \left\{
  \begin{array}{ll}
  1, &\text{if}~ (\gamma_1, \gamma_2)\in\mathbf R_0, \\
  \frac{1}{\rho}(\frac{1}{\eta_1}{\phi_1^{2,2}} + \frac{1}{\eta_2}{\phi_2^{2,2}} ), &\text{if}~ (\gamma_1, \gamma_2)\in \mathbf S_1,\\
  \frac{1}{\rho}(\frac{1}{\eta_1}{\phi_1^{1,1}} + \frac{1}{\eta_2}{\phi_2^{2,2}}), &\text{if}~ (\gamma_1, \gamma_2)\in \mathbf S_2,\\
  \frac{1}{\rho}(\frac{1}{\eta_1} {\phi_1^{1,2}} + \frac{1}{\eta_2} {\phi_2^{2,2}} ), &\text{if}~ (\gamma_1, \gamma_2)\in \mathbf S_3,
  \end{array}
  \right.
\end{align}
where the notation `$\sim$' on a probability is used to denote the asymptotic result in the high-SNR regime. On the other hand, with regard to the outage performance of the LUF strategy, by applying Corollary \ref{Coro_HUF} and Corollary \ref{Coro_LUF} we have
${{\mathcal P}}^{\mathcal L}_{\text{sys}}={{\mathcal P}}^{\mathcal H}_{\text{sys}}$, and correspondingly $\tilde {{\mathcal P}}^{\mathcal L}_{\text{sys}}=\tilde {{\mathcal P}}^{\mathcal H}_{\text{sys}}$.

\begin{figure}[!t]
  \centering
  \includegraphics[width=2.8in]{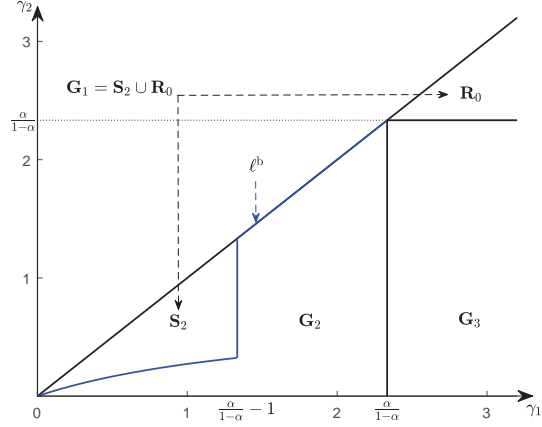}
  \caption{An illustration of $\mathbf G_k$, $k=1,2,3$, with $\alpha=0.7$.}\label{fig_gain_regions}\vspace{-5mm}
\end{figure}

To show the effectiveness of the proposed strategies, we next compare the system outage probability achieved by the proposed strategies with the counterpart achieved by the most widely adopted CSD strategy. According to the rationale in \cite{Ding_random_users}, the PA-DOS of the CSD strategy can be written as $[\pi, \omega_1, \omega_2]=[2,2,2]$ if $|h_1|^2\geq|h_2|^2$, and $[\pi, \omega_1, \omega_2]=[1,1,1]$ otherwise. Using (\ref{Form_Psys}), the system outage probability achieved by the CSD strategy can be written as
\begin{align}
  {\mathcal P}^{\mathcal C}_{\text{sys}}
  = 1-\mathcal P\{\mathbb S_1^{2,2},\mathbb S_2^{2,2}, |h_1|^2\geq|h_2|^2\}-\mathcal P\{\mathbb S_1^{1,1},\mathbb S_2^{1,1},|h_1|^2<|h_2|^2\}.
\end{align}
Recall that, when $(\gamma_1,\gamma_2)\in\mathbf R_0$, $\mathbb S_1^{\pi,\omega_1}=\mathbb S_2^{\pi,\omega_2}=\mathbb I$ for any $[\pi,\omega_1,\omega_2]$ due to $\mathbf P=\emptyset$. Moreover, $[2,2,2]$ always satisfies $2\in\mathbf P$ and $2\in\mathbf O(2)$ when $(\gamma_1,\gamma_2)\in\mathbf R^{\text{ava}}$, whereas $[1,1,1]$ satisfies $1\in\mathbf P$ and $1\in\mathbf O(1)$ only when $(\gamma_1, \gamma_2)\in \cup_{k=1,2,4}\mathbf R_k$.
Applying these facts with (\ref{Event_Si}) and following the similar steps from (\ref{Pout_HUF_sys_formu}) to (\ref{Pouts_HUF3}), ${\mathcal P}^{\mathcal C}_{\text{sys}}$ can be calculated as
\begin{align}\label{Pout_CSD_sys}
  &{\mathcal P}^{\mathcal C}_{\text{sys}}
   =\left\{
    \begin{array}{ll}
    1, &\text{if}~(\gamma_1, \gamma_2)\in \mathbf R_0, \\
    1- e^{-\frac{1}{\rho} (\frac{1}{\eta_1} {\phi_1^{2,2}} + \frac{1}{\eta_2} {\phi_2^{2,2}} )}
     + \tfrac{{1}/{\eta_1}}{{1}/{\eta_1}+{1}/{\eta_2}} e^{-\frac{1}{\rho} {\phi_1^{2,2}} (\frac{1}{\eta_1}+\frac{1}{\eta_2})}, &\text{if}~(\gamma_1, \gamma_2)\in \cup_{k=3,5}\mathbf R_k, \\
    1- e^{-\frac{1}{\rho} (\frac{1}{\eta_1} {\phi_1^{2,2}} + \frac{1}{\eta_2} {\phi_2^{2,2}} )}
     - e^{-\frac{1}{\rho} (\frac{1}{\eta_1} {\phi_1^{1,1}} + \frac{1}{\eta_2} {\phi_2^{1,1}} )} \\
     + \tfrac{{1}/{\eta_1}}{{1}/{\eta_1}+{1}/{\eta_2}} e^{-\frac{1}{\rho} {\phi_1^{2,2}} (\frac{1}{\eta_1}+\frac{1}{\eta_2})}
     + \tfrac{{1}/{\eta_2}}{{1}/{\eta_1}+{1}/{\eta_2}} e^{-\frac{1}{\rho} {\phi_2^{1,1}} (\frac{1}{\eta_1}+\frac{1}{\eta_2})}, &\text{if}~(\gamma_1, \gamma_2)\in \cup_{k=1,2,4}\mathbf R_k.
    \end{array}
    \right.
\end{align}
Employing $1-e^{-x}\overset{x\rightarrow 0}{\simeq} x$ in (\ref{Pout_CSD_sys}), the high-SNR asymptotic result of ${{\mathcal P}}^{\mathcal C}_{\text{sys}}$ is
\begin{align}\label{Pout_CSD_sys_asymp}
  \tilde {{\mathcal P}}^{\mathcal C}_{\text{sys}}=
    \left\{
    \begin{array}{ll}
    1, &\text{if}~ (\gamma_1, \gamma_2)\in\mathbf R_0, \\
    \frac{1/\eta_1}{1/\eta_1+1/\eta_2}, &\text{if}~ (\gamma_1, \gamma_2)\in \cup_{k=3,5}\mathbf R_k,\\
    \frac{1}{\rho}(\frac{1}{\eta_1}{\phi_1^{1,1}} + \frac{1}{\eta_2}{\phi_2^{2,2}}), &\text{if}~ (\gamma_1, \gamma_2)\in \cup_{k=1,2,4}\mathbf R_k.
    \end{array}
  \right.
\end{align}
Further, for the tractability of the comparison, we define the following coding gain:
\begin{align}\label{Def_Gain_CSD}
  G \triangleq \lim_{\rho\rightarrow\infty} 10\log_{10} \frac{{\mathcal P}^{\mathcal C}_{\text{sys}}}{{\mathcal P}^{\mathcal H}_{\text{sys}}},
\end{align}
which is used to measure how many times ${\mathcal P}^{\mathcal H}_{\text{sys}}$ ($={\mathcal P}^{\mathcal L}_{\text{sys}}$) is smaller than ${\mathcal P}^{\mathcal C}_{\text{sys}}$ in dB within the high-SNR regime.
Since $\log x$ is continuous in $(0,\infty)$, according to the limits of compositions we have $\lim_{\rho\rightarrow\infty} 10\log {{\mathcal P}^{\mathcal C}_{\text{sys}}}/{{\mathcal P}^{\mathcal H}_{\text{sys}}}
=10\log \lim_{\rho\rightarrow\infty} {{\mathcal P}^{\mathcal C}_{\text{sys}}}/{{\mathcal P}^{\mathcal H}_{\text{sys}}}$.
Applying this fact and $\lim_{x\rightarrow 0} \frac{e^{x}-1}{x}=\lim_{x\rightarrow 0} \frac{x}{x}=1$, while utilizing the results in  (\ref{Pout_HUF_asym}) and (\ref{Pout_CSD_sys_asymp}), we obtain
\begin{align}\label{Gain_CSD}
   G
  =\left\{
  \begin{array}{ll}
  0, &\text{if}~ (\gamma_1, \gamma_2)\in \mathbf G_1\triangleq  \mathbf S_2\cup\mathbf R_0, \\
  \theta, &\text{if}~ (\gamma_1, \gamma_2)\in \mathbf G_2\triangleq (\mathbf S_1\cup\mathbf S_3) \cap \cup_{k=1,2,4}\mathbf R_k, \\
  \infty, &\text{if}~ (\gamma_1, \gamma_2)\in \mathbf G_3\triangleq  \cup_{k=3,5}\mathbf R_k,
  \end{array}
  \right.
\end{align}
where $\mathbf G_1$, $\mathbf G_2$, and $\mathbf G_3$ are illustrated by Fig.  \ref{fig_gain_regions}, and $\theta$ is a constant unrelated to $\rho$, which is defined as $\theta= 10\log(\frac{1}{\eta_1}\phi_1^{1,1} + \frac{1}{\eta_2}\phi_2^{2,2})/(\frac{1}{\eta_1}\phi_1^{2,2} + \frac{1}{\eta_2}\phi_2^{2,2})$ if $(\gamma_1, \gamma_2)\in \mathbf S_1\cap \cup_{k=1,2,4}\mathbf R_k$, and
$\theta= 10\log(\frac{1}{\eta_1}\phi_1^{1,1} + \frac{1}{\eta_2}\phi_2^{2,2})/(\frac{1}{\eta_1} \phi_1^{1,2} + \frac{1}{\eta_2} \phi_2^{2,2})$ if $(\gamma_1, \gamma_2)\in \mathbf S_3\cap \cup_{k=1,2,4}\mathbf R_k$.

As revealed by (\ref{Gain_CSD}), when $(\gamma_1,\gamma_2)\in\mathbf G_1$, the CSD strategy can perform as well as the proposed strategies in the high-SNR regime. However, when $(\gamma_1, \gamma_2)\in\mathbf G_2$, the proposed strategies achieve a coding gain equal to $\theta$, on which we have the following observations. According to (\ref{O_iprime_P1}) and (\ref{Ineq_phii11_phii22}), we have $\theta\geq 0$, and the equality holds only when $(\gamma_1,\gamma_2)$ locates at the left boundary of $\mathbf G_2$, which can be expressed as $\ell^{\text{b}}=\cup_{k=1,2,3} \ell^{\text{b}}_k$ with
\begin{align}\label{left_bound}
  \left\{
  \begin{array}{l}
  \ell^{\text{b}}_1= \{(\gamma_1, \gamma_2): \frac{\alpha}{1-\alpha}>\gamma_1>\frac{\alpha}{1-\alpha}-1, \gamma_2=\gamma_1 \}, \\
  \ell^{\text{b}}_2= \{(\gamma_1, \gamma_2): \gamma_1=\frac{\alpha}{1-\alpha}-1, \frac{\alpha}{1-\alpha}-1\geq \gamma_2>1/(\frac{1-\alpha}{2\alpha-1}+\frac{\alpha}{1-\alpha}) \}, \\
  \ell^{\text{b}}_3= \{(\gamma_1, \gamma_2): \frac{\alpha}{1-\alpha}-1\geq\gamma_1>0, \gamma_2=\frac{(1-\alpha)\gamma_1}{\alpha+(2\alpha-1)\gamma_1} \}.
  \end{array}
  \right.
\end{align}
On the other hand, as can be verified by using the partial derivation, $\theta$ monotonically increases along with the increase of $\gamma_1$ and $d_1$, or the decrease of $\gamma_2$ and $d_2$.
At last, when $(\gamma_1, \gamma_2)\in\mathbf G_3$, the coding gain goes to the infinite. This is because the system outage probability achieved by CSD reaches an error floor in the high-SNR regime as shown by (\ref{Pout_CSD_sys_asymp}), whereas the system outage probability achieved by the proposed strategies keeps decreasing at the linear speed of $\rho$.

\subsection{User Outage Probabilities Achieved by the Proposed Strategies}
Recall that, when $(\gamma_1,\gamma_2)\in\mathbf R_0$, the user outage inevitably occurs for both $U_1$ and $U_2$, and when $(\gamma_1,\gamma_2)\in\mathbf R^{\text{ava}}$, both HUF and LUF guarantee $\pi\in\mathbf P$ and $\omega_1,\omega_2\in\mathbf O(\pi)$. Applying these facts in (\ref{Form_Pi}) and (\ref{Event_Si}), the user outage probability achieved by HUF and LUF can be written as $\mathcal P_i^{\mathcal H}=\mathcal P\{ \rho|h_i|<\phi_i^{\pi^{\mathcal H},\omega_i^{\mathcal H}} \}$ and $\mathcal P_i^{\mathcal L}=\mathcal P\{ \rho|h_i|<\phi_i^{\pi^{\mathcal L},\omega_i^{\mathcal L}} \}$ for $i=1,2$. Then, following the similar steps from (\ref{Pout_HUF_sys_formu}) to (\ref{Pouts_HUF3}), $\mathcal P_i^{\mathcal H}$ and $\mathcal P_i^{\mathcal L}$ can be readily obtained as follows:
When $(\gamma_1, \gamma_2)\in\mathbf R_0$,
\begin{align}\label{Pi_A0}
  {\mathcal P}^{\mathcal H}_i={\mathcal P}^{\mathcal L}_i=1, ~\text{for}~i=1,2;
\end{align}
When $(\gamma_1, \gamma_2)\in \mathbf S_1$,
\begin{align}
  {\mathcal P}^{\mathcal H}_i={\mathcal P}^{\mathcal L}_i
  =1- e^{-\frac{1}{\rho\eta_i}{\phi_i^{2,2}} },  ~\text{for}~i=1,2;
\end{align}
When $(\gamma_1, \gamma_2)\in \mathbf S_2$,
\begin{align}
 \left\{
 \begin{array}{ll}
  {\mathcal P}^{\mathcal H}_1=1- e^{-\frac{1}{\rho\eta_1}{\phi_1^{1,1}} },\\
  {\mathcal P}^{\mathcal H}_2=\chi(\mathbf a_1),
 \end{array}
 \right.
 \text{and}~
 \left\{
 \begin{array}{ll}
  {\mathcal P}^{\mathcal L}_1=\chi(\mathbf a_2), \\
  {\mathcal P}^{\mathcal L}_2=1- e^{-\frac{1}{\rho\eta_2}{\phi_2^{2,2}} };
 \end{array}
 \right.
\end{align}
When $(\gamma_1, \gamma_2)\in \mathbf S_3$,
\begin{align}\label{Pi_S3}
 \left\{
 \begin{array}{ll}
  {\mathcal P}^{\mathcal H}_1=1- e^{-\frac{1}{\rho\eta_1}{\phi_1^{1,2}} },\\
  {\mathcal P}^{\mathcal H}_2=\chi(\mathbf a_3),
 \end{array}
 \right.
 \text{and}~
 \left\{
 \begin{array}{ll}
  {\mathcal P}^{\mathcal L}_1=\chi(\mathbf a_4),\\
  {\mathcal P}^{\mathcal L}_2=1- e^{-\frac{1}{\rho\eta_2}{\phi_2^{2,2}} },
 \end{array}
 \right.
\end{align}
where $\chi(\cdot)$ is defined as
$\chi(x,y,z)\triangleq 1-\exp(-\frac{1}{\rho}x)-\exp(-\frac{1}{\rho}(y+z))+\exp(-\frac{1}{\rho}(x+y))$, vectors $\mathbf a_k$, $k=1,2,3,4$, are defined as
$\mathbf a_1 \triangleq (\frac{1}{\eta_2}\phi_2^{1,1},\frac{1}{\eta_1}\phi_1^{2,2},\frac{1}{\eta_2}\phi_2^{2,2})$,
$\mathbf a_2\triangleq (\frac{1}{\eta_1}\phi_1^{2,2},\frac{1}{\eta_2}\phi_2^{1,1},\frac{1}{\eta_1}\phi_1^{1,1})$,
$\mathbf a_3 \triangleq (\frac{1}{\eta_2}\phi_2^{1,2},\frac{1}{\eta_1}\phi_1^{2,2},\frac{1}{\eta_2}\phi_2^{2,2})$,
and $\mathbf a_4\triangleq (\frac{1}{\eta_1}\phi_1^{2,2},\frac{1}{\eta_2}\phi_2^{1,2},\frac{1}{\eta_1}\phi_1^{1,2})$.

Moreover, according to Corollary \ref{Coro_HUF} and Corollary \ref{Coro_LUF} we have ${\mathcal P}^{\text{min}}_1={\mathcal P}^{\mathcal H}_1$ and ${\mathcal P}^{\text{min}}_2={\mathcal P}^{\mathcal L}_2$. Applying $1-e^{-x}\overset{x\rightarrow 0}{\simeq} x$ into the above results, the high-SNR asymptotic user outage probabilities can be obtained as
\begin{align}\label{P1_asym}
  \tilde {\mathcal P}^{\mathcal H}_1=\tilde {\mathcal P}^{\mathcal L}_1=\tilde {{\mathcal P}}^{\text{min}}_1=
  \left\{
  \begin{array}{ll}
  1, &\text{if}~ (\gamma_1, \gamma_2)\in\mathbf R_0, \\
  \frac{1}{\rho\eta_1} {\phi_1^{2,2}}, &\text{if}~ (\gamma_1, \gamma_2)\in \mathbf S_1,\\
  \frac{1}{\rho\eta_1} {\phi_1^{1,1}}, &\text{if}~ (\gamma_1, \gamma_2)\in \mathbf S_2,\\
  \frac{1}{\rho\eta_1} {\phi_1^{1,2}}, &\text{if}~ (\gamma_1, \gamma_2)\in \mathbf S_3,
  \end{array}
  \right.
\end{align}\vspace{-1mm}
and
\begin{align}\label{P2_asym}
  \tilde {\mathcal P}^{\mathcal H}_2=\tilde {\mathcal P}^{\mathcal L}_2=\tilde {{\mathcal P}}^{\text{min}}_2=
  \left\{
  \begin{array}{ll}
  1, &\text{if}~ (\gamma_1, \gamma_2)\in\mathbf R_0, \\
  \frac{1}{\rho\eta_2} {\phi_2^{2,2}}, &\text{otherwise}.
  \end{array}
  \right.
\end{align}

Recall that, when $(\gamma_1, \gamma_2)\in \mathbf S_2\cup\mathbf S_3$, ${\mathcal P}^{\text{min}}_1$ and ${\mathcal P}^{\text{min}}_2$ cannot be simultaneously achieved by any PA-DOS strategy due to the power contradiction, and thus HUF and LUF are designed to achieve the minimum user outage probability for the high-priority user in this case. However, (\ref{P1_asym}) and (\ref{P2_asym}) show an interesting phenomenon that, $\tilde {\mathcal P}^{\text{min}}_1$ and $\tilde {\mathcal P}^{\text{min}}_2$ can be simultaneously approached in the high-SNR regime by either HUF or LUF. Namely, when the transmitting SNR is sufficiently large, the low-priority user can also achieve a great outage performance with negligible gap to its optimal, indicating that the impact of the power contradiction in F-NOMA is almost eliminated.

\section{Numerical Results}\label{Section_simulation}
\begin{figure}[!t]
  \begin{minipage}[t]{0.46\textwidth}
    \centering
    \includegraphics[width=2.5in]{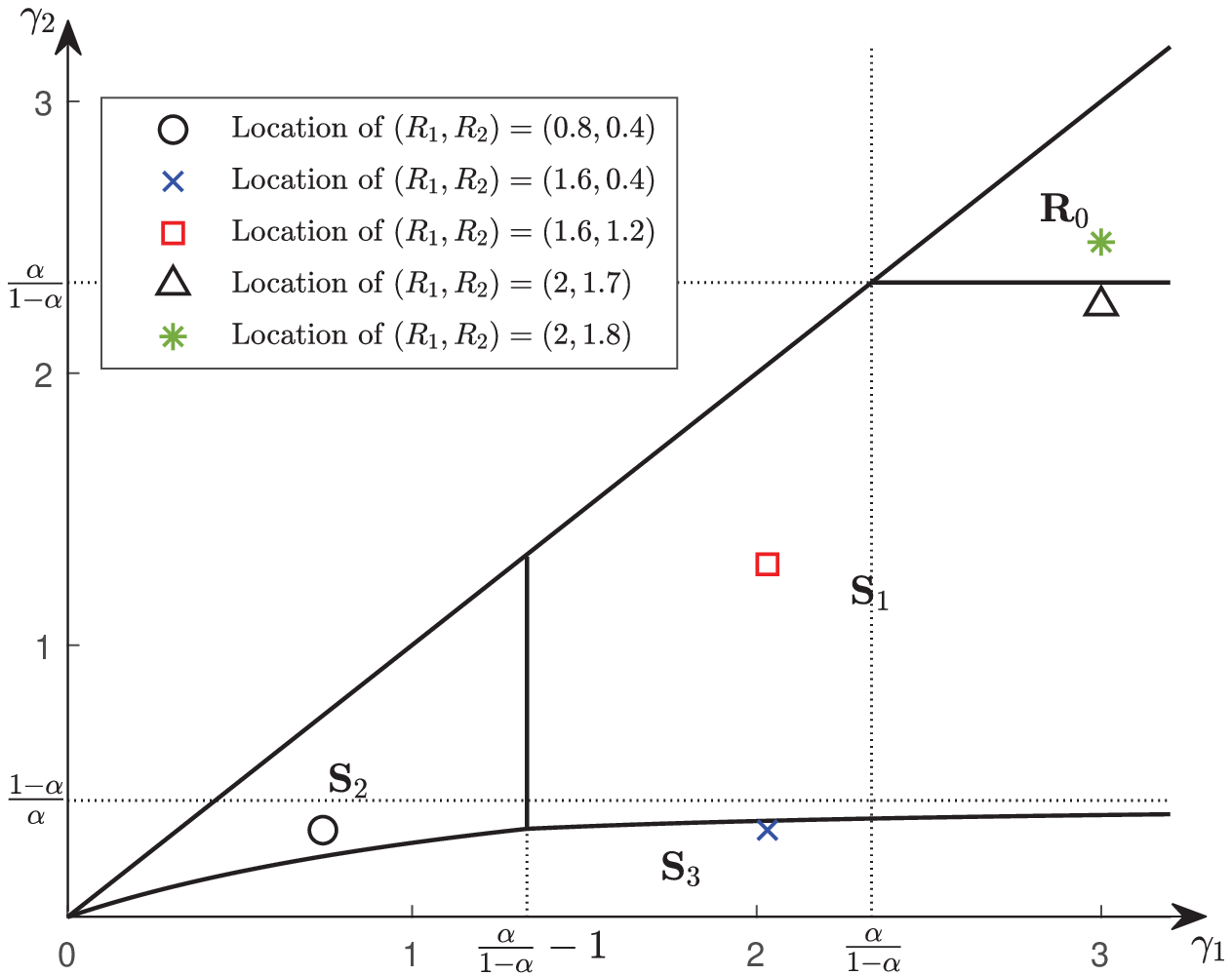}
    \caption{The corresponding locations of the considered target rate pairs in the region of decoding thresholds.}\label{fig_position}\vspace{-5mm}
  \end{minipage}
  \quad
  \begin{minipage}[t]{0.5\textwidth}
    \centering
    \includegraphics[width=2.7in]{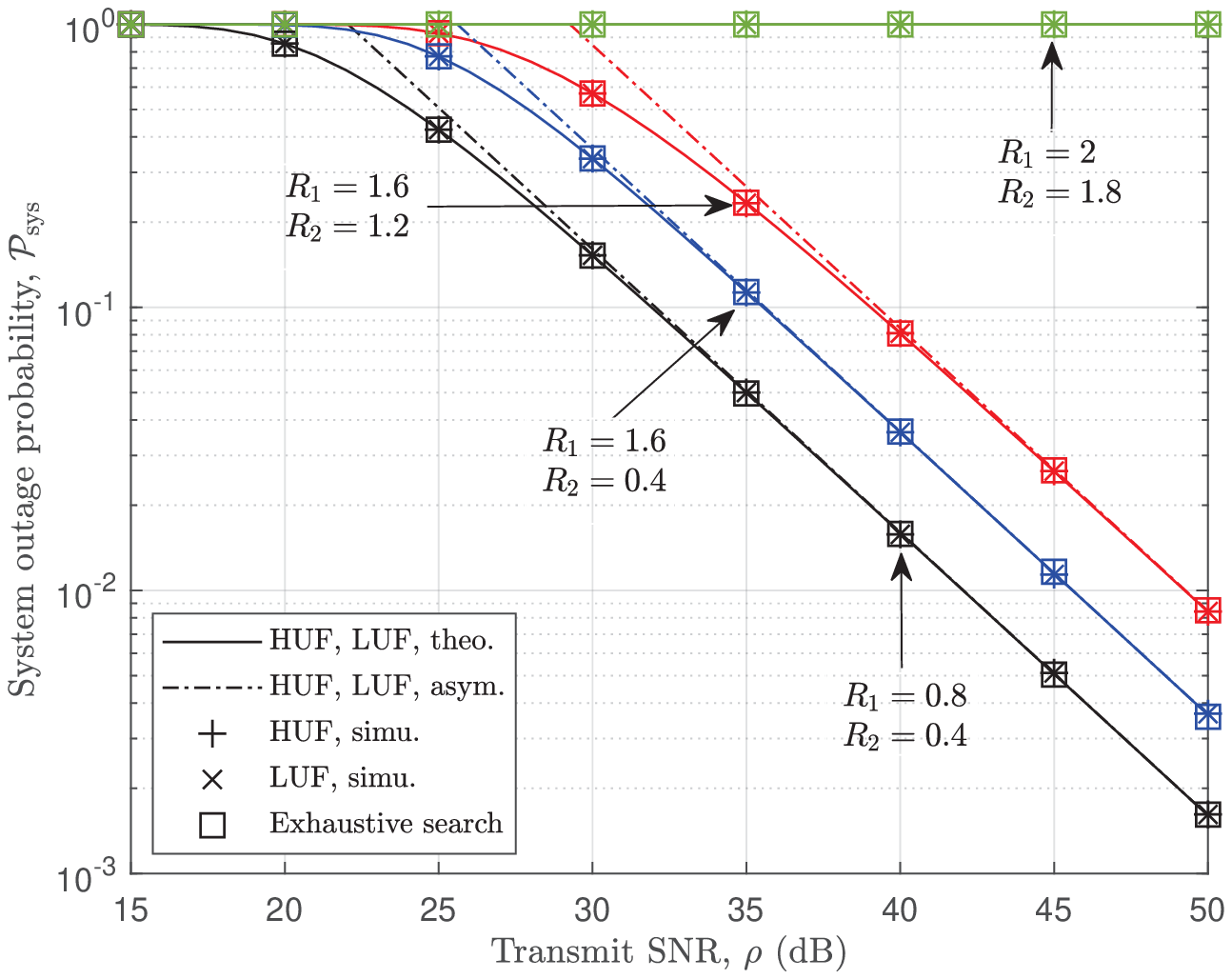}
    \caption{Simulated and theoretical system outage probabilities achieved by the proposed strategies with $d_1=d_2=40 \ m$ and $(R_1,R_2)=(0.8,0.4), (1.6,0.4), (1.6,1.2), (2,1.8)$ bps/Hz.}\label{fig_verification}\vspace{-5mm}
  \end{minipage}
\end{figure}

In the simulation, the reference distance, pathloss exponent, and the power coefficient are set to $d_0=10 \ m$, $\nu=2.7$, and $\alpha=0.7$, respectively. Five pairs of user's target rates are used for the simulation, which are $(R_1,R_2)=(0.8,0.4), (1.6,0.4), (1.6,1.2), (2.1,7), (2,1.8)$ bit/Hz. By applying $\gamma_i=2^{R_i}-1$ for $i=1,2$, Fig. \ref{fig_position} plots the locations where these five pairs of target rates are mapped in the region of $(\gamma_1,\gamma_2)$.


Fig. \ref{fig_verification} plots the simulated and theoretical system outage probabilities achieved by the proposed HUF and LUF strategies with $d_1=d_2=40 \ m$.
As can be seen from the figure, the simulated results well coincide with the theoretical results, which verifies the correctness of the derived system outage probabilities, and demonstrates that the same system outage performance is achieved by the proposed strategies. It is also shown that, the system outage probabilities under the proposed strategies coincide with the results of exhaustive search, which goes through all the combinations of $\pi$, $\omega_1$, and $\omega_2$ to avoid the system outage. This demonstrates that, the proposed strategies achieve the minimum system outage probability.



\begin{figure}[!t]
  \begin{minipage}[t]{0.46\textwidth}
    \centering
    \includegraphics[width=2.7in]{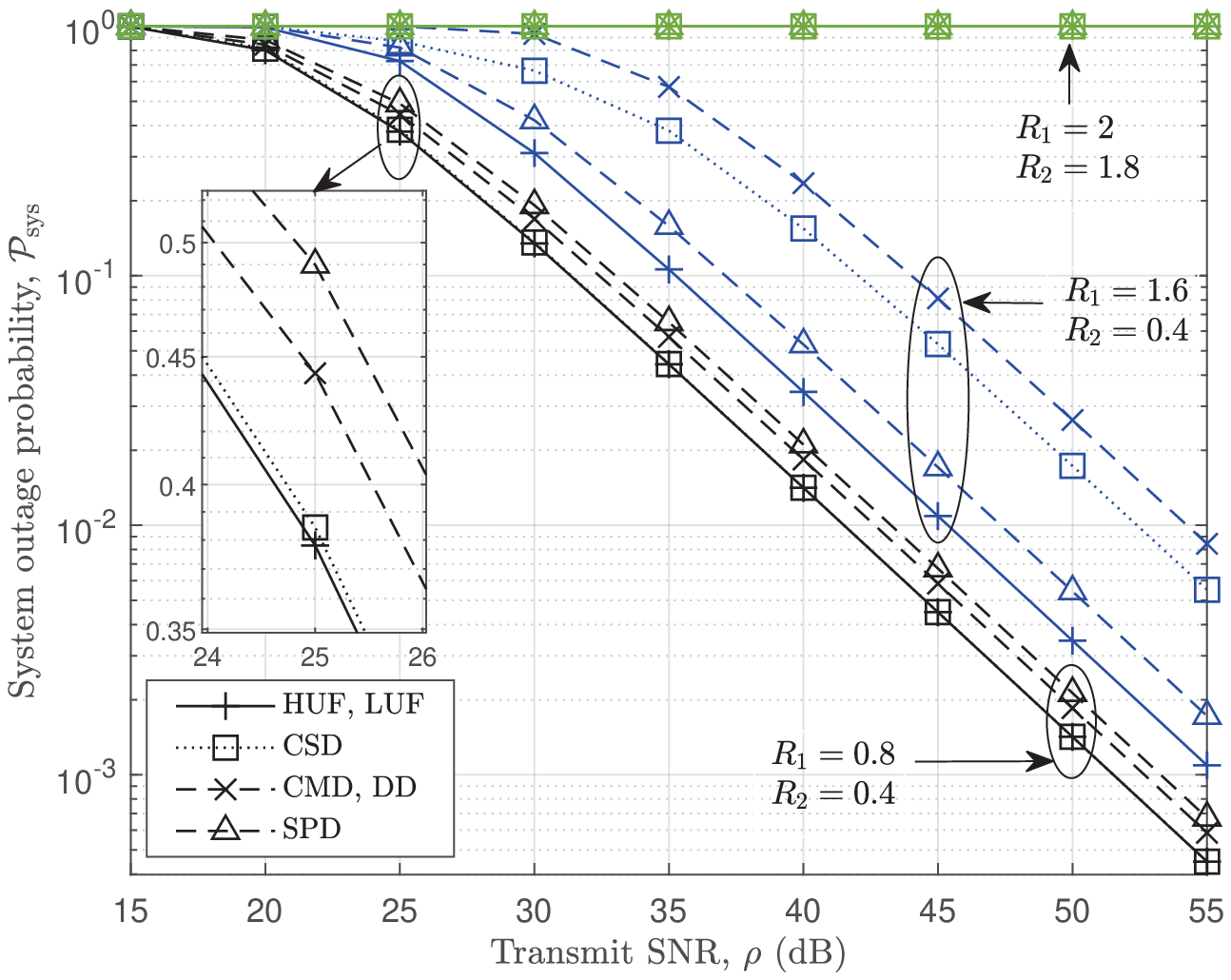}
    \caption{System outage probabilities achieved by various strategies with $d_1=40\ m$, $d_2=30\ m$, and $(R_1,R_2)=(0.8,0.4), (1.6,0.4), (2,1.8)$ bps/Hz.}\label{compare_lr_4030}\vspace{-5mm}
  \end{minipage}
  \quad
  \begin{minipage}[t]{0.5\textwidth}
    \centering
    \includegraphics[width=2.7in]{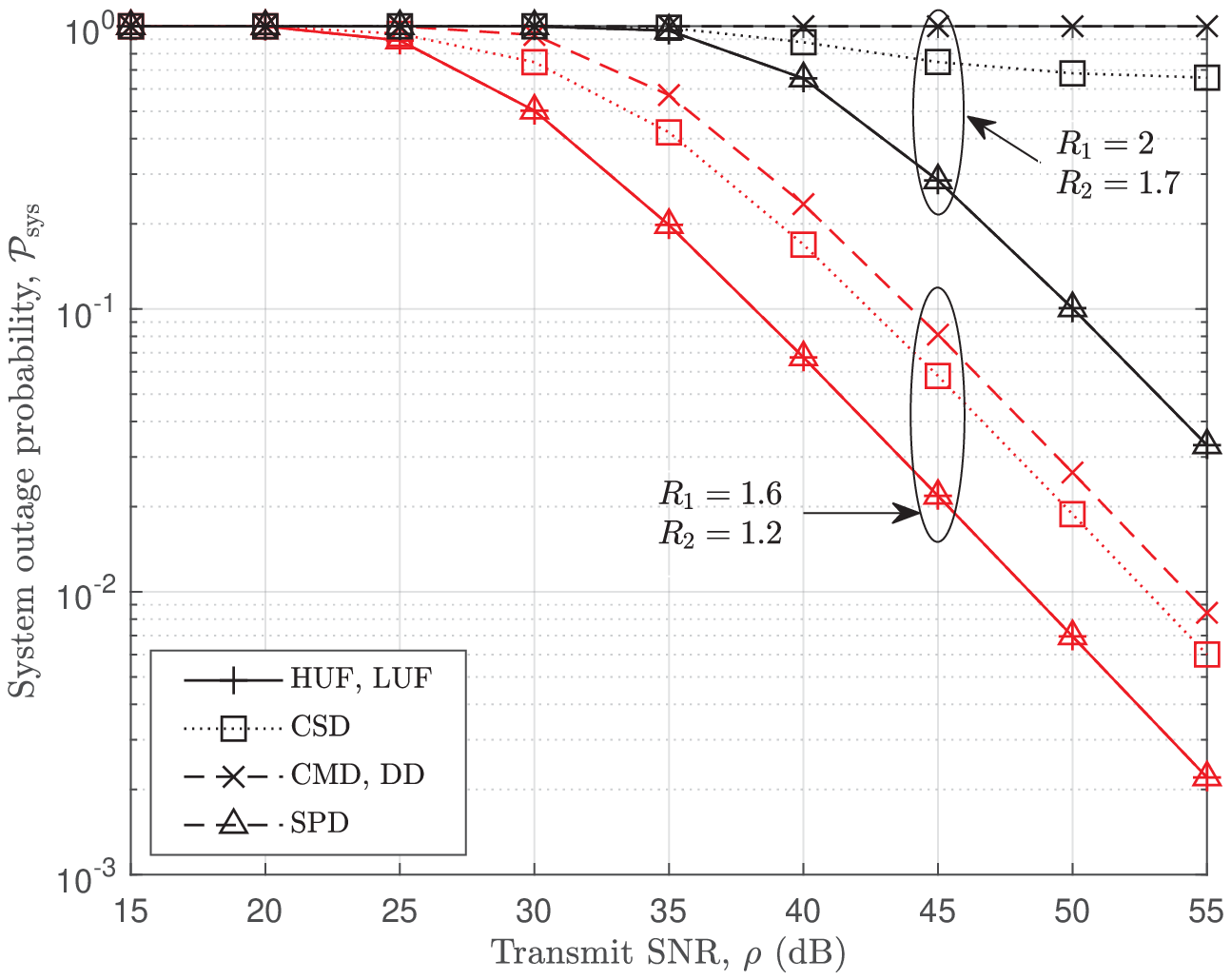}
    \caption{System outage probabilities achieved by various strategies with $d_1=40\ m$, $d_2=30\ m$, and $(R_1,R_2)=(1.6,1.2), (2,1.7)$ bps/Hz.}\label{compare_hr_4030}\vspace{-5mm}
  \end{minipage}
\end{figure}

For comparison, Fig. \ref{compare_lr_4030} plots the system outage probabilities achieved by the conventional CSD, CMD, DD, and SPD strategies as well as the proposed strategies, where $d_1=40 \ m$, $d_2=30 \ m$, and $(R_1,R_2)=(0.8,0.4), (1.6,0.4), (2,1.8)$ bps/Hz. Recall that, the CSD, CMD, and DD strategies have two possible results for the PA-DOS, i.e., $[2,2,2]$ and  $[1,1,1]$, while the criteria for these strategies to select $[2,2,2]$ instead of $[1,1,1]$ are $|h_2|^2\leq |h_1|^2$ \cite{Ding_random_users}, $\eta_2\leq\eta_1$ \cite{Kim_CMD}, and $d_2\geq d_1$ \cite{GX_DD}, respectively. On the other hand, the SPD strategy in \cite{Ding_QoS,LY_QoS,Lulv_QoS} assumes the low-rate user to have the higher service priority, and it always selects $[2,2,2]$.
As can be seen from Fig. \ref{compare_lr_4030}, when $(R_1,R_2)=(2,1.8)$ bit/Hz (satisfying $(\gamma_1,\gamma_2)\in\mathbf R_0$), all the strategies lead to a system outage probability equal to $1$, since the F-NOMA operation condition is violated.
When $(R_1,R_2)=(0.8,0.4)$ (satisfying $(\gamma_1,\gamma_2)\in\mathbf S_2$), the CMD, DD, and SPD strategies achieve higher system outage probabilities than the proposed strategies, while the CSD strategy behaves the similar outage performance with the proposed ones. When $(R_1,R_2)=(1.6,0.4)$ (satisfying $(\gamma_1,\gamma_2)\in\mathbf S_3$), the proposed strategies achieve a significantly lower system outage probability than all the other strategies.

The comparison is further illustrated in Fig. \ref{compare_hr_4030}, where $d_1=40 \ m$, $d_2=30 \ m$, and $(R_1,R_2)=(1.6,1.2),(2,1.7)$ bps/Hz (satisfying $(\gamma_1,\gamma_2)\in\mathbf S_1$). It can be observed that, for both of the two rate pairs, the SPD strategy achieves the same outage performance as the proposed strategies, since they make the same decision on PA-DOS when $(\gamma_1,\gamma_2)\in\mathbf S_1$. In contrast, the CSD, CMD, and DD strategies achieve much higher system outage probabilities. Especially when $(R_1,R_2)=(2,1.7)$, the system outage probability achieved by CSD, CMD, and DD reach error floors, which are avoided under the proposed strategies by properly performing PA-DOS.

\begin{figure}[!t]
\centering
  \subfigure[$G$ versus $\gamma_1$ and $\gamma_2$ with $d_1=d_2=40 \ m$]{\label{fig_gainCSD_gamma} 
  \includegraphics[width=2.7in]{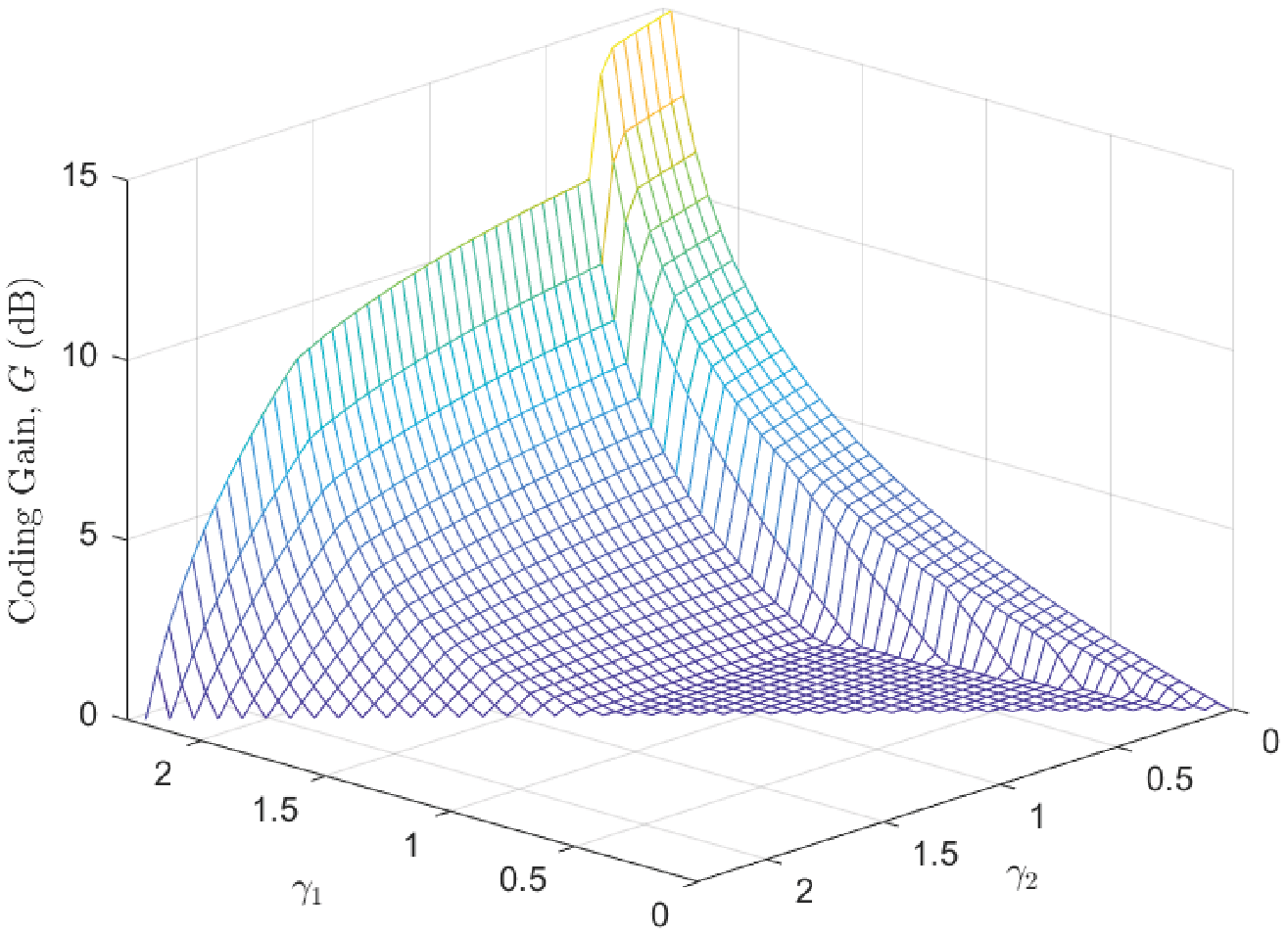}}
  \quad\quad
  \subfigure[$G$ versus $d_1$ and $d_2$ with $(R_1,R_2)=(1.6,1.2)$ bit/Hz]{ \label{fig_gainCSD_dis} 
  \includegraphics[width=2.7in]{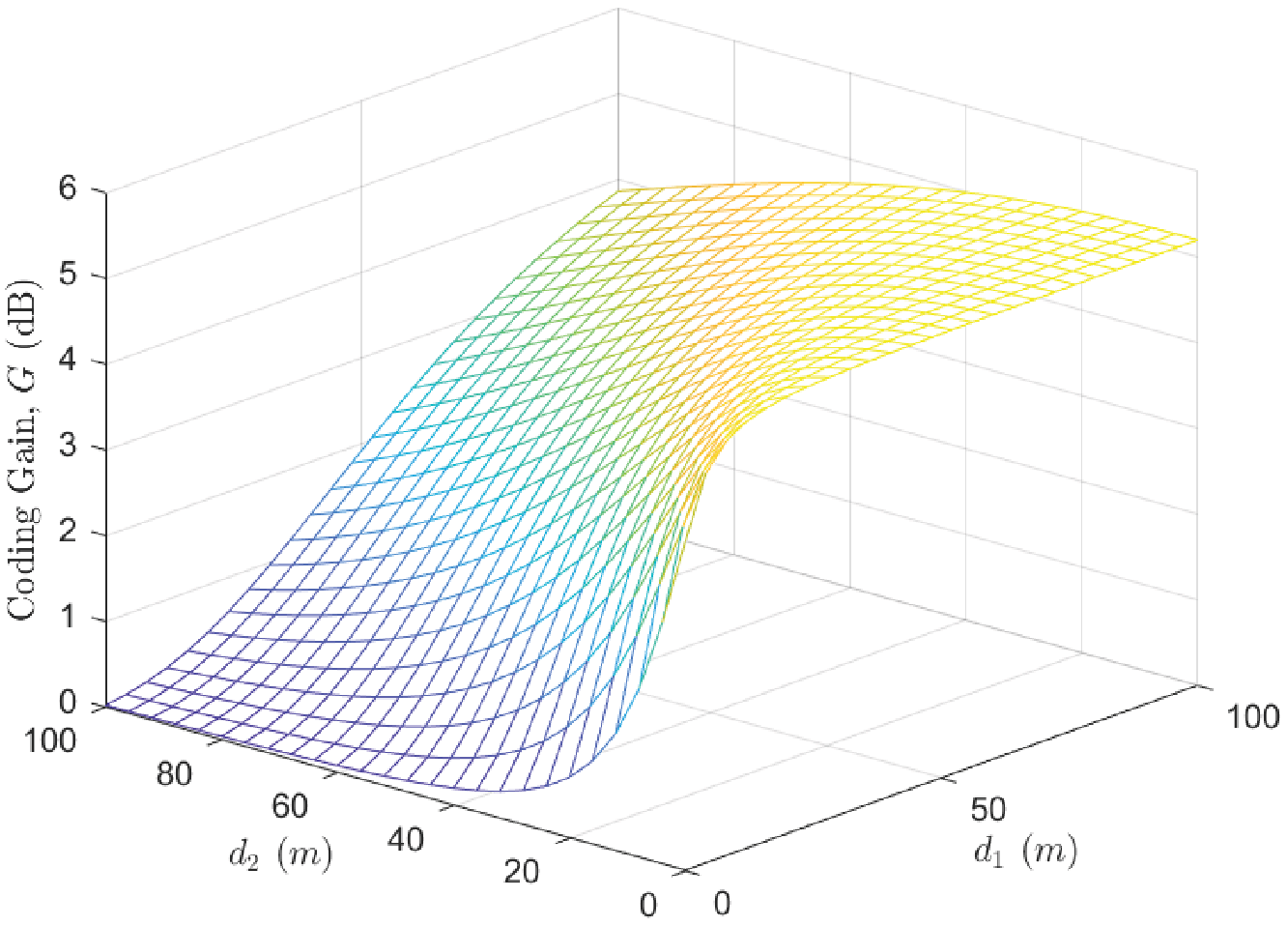}}
\caption{The coding gain of the proposed strategies on the CSD strategy.}\vspace{-7mm}
\label{fig_gainCSD}
\end{figure}

Fig. \ref{fig_gainCSD} depicts the coding gain of the proposed strategies on the CSD strategy, which is given by (\ref{Gain_CSD}). Specifically, Fig. \ref{fig_gainCSD_gamma} plots $G$ versus $\gamma_1$ and $\gamma_2$, where $(\gamma_1,\gamma_2)$ is in region $\mathbf S_2\cup\mathbf G_2$ for finite values of $G$. By comparing Fig. \ref{fig_gainCSD_gamma} with Fig. \ref{fig_position} it can be observed that, when $(\gamma_1,\gamma_2)\in\mathbf G_2\setminus \ell^{\text{b}}$, $G$ achieves positive values. Specifically, in this region $G$ increases along with an increase of $\gamma_1$ or a decrease of $\gamma_2$. Fig. \ref{fig_gainCSD_dis} shows the impact of $d_1$ and $d_2$ on $G$ with $(R_1,R_2)=(1.6,1.2)$ bit/Hz. As can be seen, $G$ increases along with an increase of $d_1$ or a decrease of $d_2$, which is consistent with the analysis in Section \ref{subsection_Psys}.

\begin{figure}[!t]
\centering
  \subfigure[$(R_1,R_2)=(1.6,1.2)$ bit/Hz]{\label{fig_user_1612} 
  \includegraphics[width=2.7in]{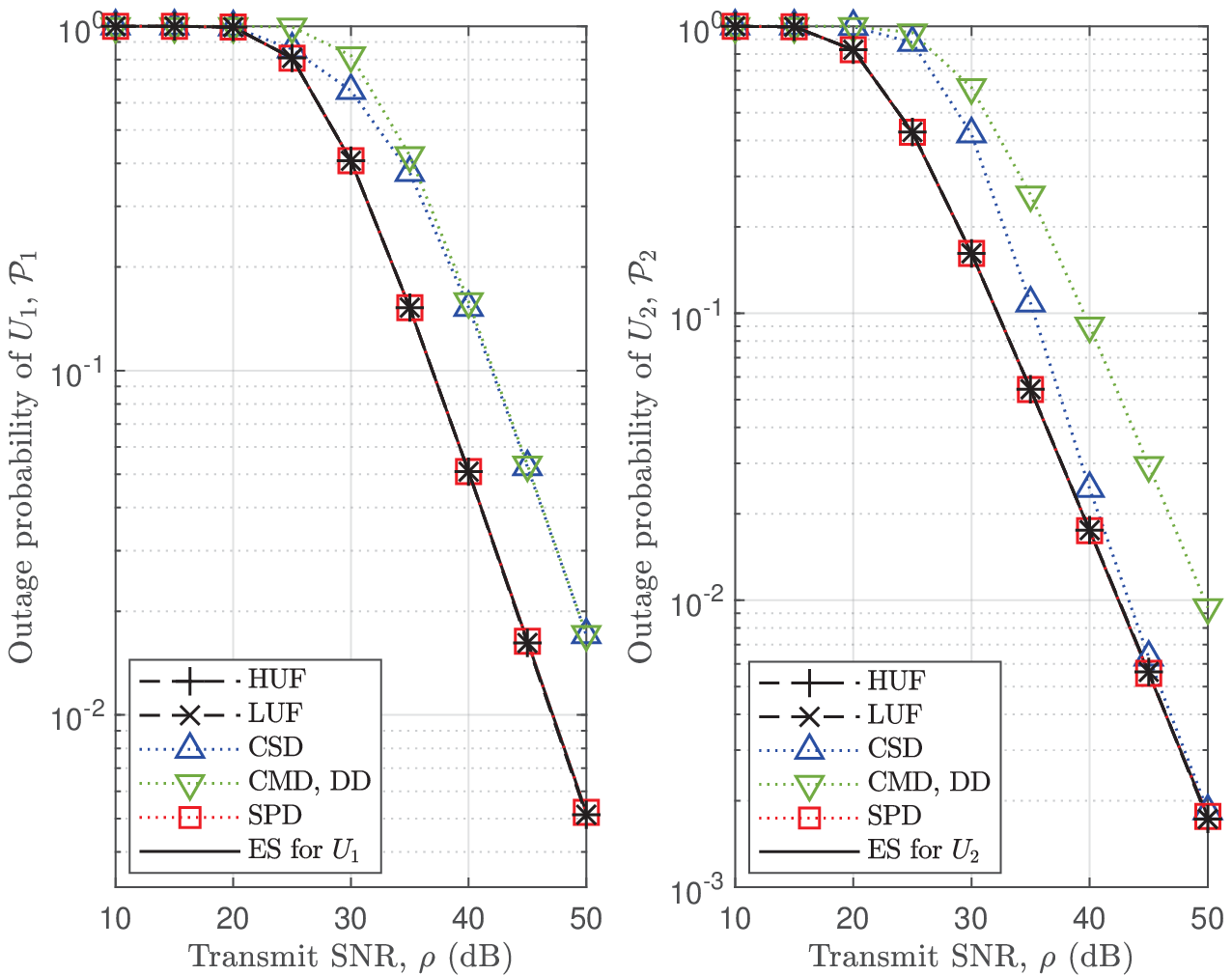}}
  \quad\quad
  \subfigure[$(R_1,R_2)=(1.6,0.4)$ bit/Hz]{ \label{fig_user_1604} 
  \includegraphics[width=2.7in]{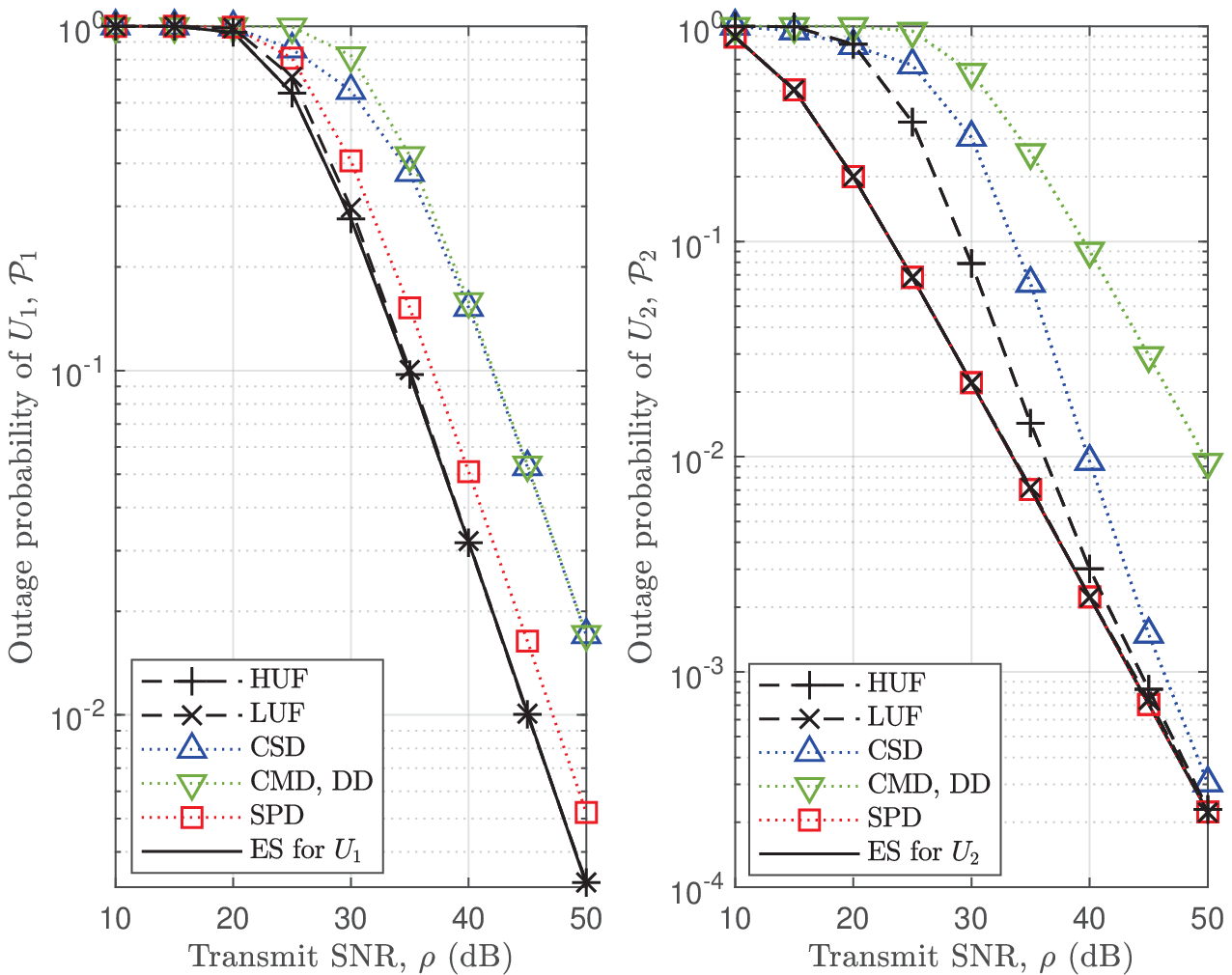}}
\caption{User outage probabilities achieved by various strategies, where $d_1=40 \ m$ and $d_2=30 \ m$.}\vspace{-6mm}
\label{fig_user}
\end{figure}

User outage probabilities are plotted in Fig. \ref{fig_user}, where $d_1=40 \ m$, $d_2=30 \ m$, and $(R_1,R_2)=(1.6,1.2),(1.6,0.4)$.
In Fig. \ref{fig_user}, `ES for $U_i$' denotes the exhaustive search, which goes through all the combinations of $\pi$ and $\omega_i$ to avoid $U_i$'s outage, and shows the minimum achievable user outage probability of $U_i$.
As can be seen from the figure, when $(R_1,R_2)=(1.6,1.2)$ (satisfying $(\gamma_1,\gamma_2)\in\mathbf S_1$), the proposed strategies can achieve the minimum user outage probabilities for both $U_1$ and $U_2$. When $(R_1,R_2)=(1.6,0.4)$ (satisfying $(\gamma_1,\gamma_2)\in\mathbf S_3$), the minimum user outage probability of $U_1$ is achieved by HUF, and it can also be approached by LUF if the transmit SNR is sufficiently large. The minimum user outage probability of $U_2$ is achieved by LUF, and it can be approached by HUF in the high-SNR regime.
For the SPD strategy, when $(R_1,R_2)=(1.6,1.2)$, it also achieves the minimum user outage probability for both $U_1$ and $U_2$, due to the same PA-DOS decision as the proposed in this case. When $(R_1,R_2)=(1.6,0.4)$, the SPD strategy achieves the minimum user outage probability for $U_2$ as LUF does, while it achieves a higher user outage probability for $U_1$ than both of the proposed strategies. Furthermore, compared with the proposed strategies, the CSD, CMD, and DD strategies perform worse outage performance for both $U_1$ and $U_2$ with either of the two rate pairs.

\section{Conclusion}
\label{Conclusion}
This paper has investigated PA-DOS in a two-user downlink F-NOMA network, where both the system outage performance and the user outage performance are examined. By employing the whole flexibility of PA-DOS, two rate-adaptive PA-DOS criteria have been proved, which reveal the significant impact of target rates on the outage performance of F-NOMA, and provide guidelines for the design of PA-DOS strategies. Based on the developed criteria, two outage-optimal PA-DOS strategies have been proposed, which achieve the minimum system outage probability and the minimum user outage probability for the high-rate user and the low-rate user, respectively. Especially, the proposed strategies are easy to implement since they do not require CSIT or only require one-bit CSI feedback depending on different target rates. For performance evaluation, analytical and simulated results have been provided, which confirm the superiority of the proposed strategies on outage performance over the existing strategies.

\appendices
\section{Proof of Lemma \ref{lemma_NScondition}}\label{Proof_lemma_NScondition}
 For simplicity, in this appendix we use $\Phi^{\mathcal A}$ to represent $\Phi^{\pi^{\mathcal A},\omega_1^{\mathcal A},\omega_2^{\mathcal A}}$, and use $\Phi^{\text{max}}$ to represent $\max_{\pi,\omega_1,\omega_2\in\{1,2\}}\Phi^{\pi,\omega_1,\omega_2}$. We first prove the necessity of condition $\mathcal P\{\Phi^{\mathcal A}<1,\Phi^{\text{max}}\geq 1\}=0$ to $\mathcal P_{\text{sys}}^{\mathcal A}=\mathcal P_{\text{sys}}^{\text{min}}$ by contradiction. Assume that, for $\mathcal A$ with $\mathcal P_{\text{sys}}^{\mathcal A}=\mathcal P_{\text{sys}}^{\text{min}}$, inequality $\mathcal P\{\Phi^{\mathcal A}<1,\Phi^{\text{max}}\geq 1\}>0$ holds. Then, we construct a new strategy $\mathcal A^\dagger$ as $[\pi^{\mathcal A^\dagger},\omega_1^{\mathcal A^\dagger},\omega_2^{\mathcal A^\dagger}]=[\pi^{\mathcal A},\omega_1^{\mathcal A},\omega_2^{\mathcal A}]$ if $\Phi^{\mathcal A}\geq 1$ and $[\pi^{\mathcal A^\dagger},\omega_1^{\mathcal A^\dagger},\omega_2^{\mathcal A^\dagger}]=\arg\max_{\pi,\omega_1,\omega_2\in\{0,1\}} \Phi^{\pi,\omega_1,\omega_2}$ otherwise. Applying $\mathcal A^\dagger$ in (\ref{Psys_re}), we have
 \begin{align}\label{Adagger_op_Astar}
   \mathcal P_{\text{sys}}^{\mathcal A^\dagger}
   &= \mathcal P\{ \Phi^{\mathcal A^\dagger}<1\} \notag\\
   &= \mathcal P\{ \Phi^{\mathcal A}<1, \Phi^{\mathcal A}\geq 1\} + \mathcal P\{ \Phi^{\text{max}}< 1, \Phi^{\mathcal A}< 1\} \notag\\
   &= \mathcal P\{ \Phi^{\mathcal A}< 1\}-\mathcal P\{\Phi^{\mathcal A}< 1, \Phi^{\text{max}}\geq 1\}\notag\\
   &< \mathcal P\{ \Phi^{\mathcal A}< 1\}=\mathcal P_{\text{sys}}^{\mathcal A}.
 \end{align}
 Recall $\mathcal P_{\text{sys}}^{\mathcal A}=\mathcal P_{\text{sys}}^{\text{min}}$, which is in contradiction with $\mathcal P_{\text{sys}}^{\mathcal A^\dagger}<\mathcal P_{\text{sys}}^{\mathcal A}$. Thus, the necessity of the condition is proved.

 Next, we prove the sufficiency of the condition. Assume that, one strategy $\mathcal A$ satisfies $\mathcal P\{\Phi^{\mathcal A}<1,\Phi^{\text{max}}\geq 1\}=0$. Based on this assumption, we have

 \begin{align}\label{Astar_op_Adagger}
   & \mathcal P \{\Phi^{\mathcal A}\geq 1 \cup \Phi^{\text{max}}< 1 \}=1\notag\\
   \Leftrightarrow& \mathcal P \{\Phi^{\mathcal A}\geq1\} + \mathcal P \{\Phi^{\text{max}}< 1 \}=1\notag\\
   \Leftrightarrow& \mathcal P \{\Phi^{\mathcal A}<1\} = \mathcal P \{\Phi^{\text{max}}< 1 \}
   \leq \mathcal P \{ \Phi^{\mathcal A'}< 1 \}, \forall \mathcal A'\in \mathbf A,
 \end{align}
 where the first step comes from the event exclusion, and the inequality in the second steps is ensured by $\{\Phi^{\text{max}}< 1 \}\subseteq\{ \Phi^{\mathcal A'}< 1 \}$ for any $\mathcal A'\in \mathbf A$, with $\mathbf A$ denoting the set of all the PA-DOS strategies. Using (\ref{Psys_re}), (\ref{Astar_op_Adagger}) leads to ${\mathcal P}^{\mathcal A}_{\text{sys}}\leq {\mathcal P}^{\mathcal A'}_{\text{sys}}$ for any $\mathcal A'\in \mathbf A$, which proves $\mathcal P_{\text{sys}}^{\mathcal A}=\mathcal P_{\text{sys}}^{\text{min}}$. This completes the proof of Lemma \ref{lemma_NScondition}.

\section{Proof of Lemma \ref{lemma_maxPhi}}\label{Proof_best_decision}
  Recall $\mathbf P=\emptyset$ when $(\gamma_1,\gamma_2)\in\mathbf R_0$, applying which in (\ref{Def_Phi}) $\max_{\pi,\omega_1,\omega_2\in\{1,2\}} \Phi^{\pi,\omega_1,\omega_2}=0$ is obtained. In the following, we focus on the case $(\gamma_1,\gamma_2)\in\mathbf R^{\text{ava}}$. By using (\ref{Def_Phi}) we have
  \begin{align}\label{maxmaxPhi_re1}
    \max_{\pi,\omega_1,\omega_2\in\{1,2\}} \Phi^{\pi,\omega_1,\omega_2}
    =\max_{\pi\in\mathbf P,\omega_1,\omega_2\in\mathbf O(\pi)} \Phi^{\pi,\omega_1,\omega_2}
    =\max_{\pi\in\mathbf P}\Phi^{\pi,\omega_1^\prime(\pi),\omega_2^\prime(\pi)},
  \end{align}
  where $\omega_i^\prime(\pi)\triangleq \arg\min_{\omega_i\in\mathbf O(\pi)}  \phi_i^{\pi,\omega_i}$ for any $i\in\{1,2\}$ and $\pi\in\mathbf P$. Further, applying (\ref{Set_O}), (\ref{Set_P}), (\ref{phi_1}), and (\ref{phi_2}), $\omega_i^\prime(\pi)$ can be found as, for $i=1,2$,
  \begin{align}\label{O_iprime_P1}
    \omega_i^\prime(1)
    =\left\{
    \begin{array}{ll}
    1, &\text{if}~\gamma_2\geq \frac{(1-\alpha)\gamma_1}{\alpha+(2\alpha-1)\gamma_1},\\
    2, &\text{otherwise},
    \end{array}
    \right.
  \end{align}\vspace{-8mm}
  \begin{align}\label{O_iprime_P2}
    \omega_i^\prime(2)=2,
  \end{align}
  where (\ref{O_iprime_P2}) is valid for $(\gamma_1,\gamma_2)\in\mathbf R^{\text{ava}}$, while (\ref{O_iprime_P1}) is only valid for $(\gamma_1,\gamma_2)\in\cup_{k=1}^{4}\mathbf R_k$, since $\pi=1\notin \mathbf P$ when $(\gamma_1,\gamma_2)\in\mathbf R_5$.
  Based on this result, we calculate $\max_{\pi\in\mathbf P}\Phi^{\pi,\omega_1^\prime(\pi),\omega_2^\prime(\pi)}$ under $(\gamma_1,\gamma_2)\in\cup_{k=1}^{4}\mathbf R_k$ and $(\gamma_1,\gamma_2)\in\mathbf R_5$, respectively.

  Considering $(\gamma_1,\gamma_2)\in\cup_{k=1}^{4}\mathbf R_k$, by applying (\ref{O_iprime_P1}), (\ref{O_iprime_P2}), and $\mathbf P=\{1,2\}$ we have
  \begin{align}\label{max_Phi_Oprime_1to4}
    \max_{\pi\in\mathbf P}\Phi^{\pi,\omega_1^\prime(\pi),\omega_2^\prime(\pi)}
    =\left\{
      \begin{array}{ll}
      \max \{ \Phi^{1,1,1} , \Phi^{2,2,2} \}, &\text{if}~\gamma_2\geq \frac{(1-\alpha)\gamma_1}{\alpha+(2\alpha-1)\gamma_1},\\
      \max \{ \Phi^{1,2,2} , \Phi^{2,2,2} \}, &\text{otherwise}.
      \end{array}
      \right.
  \end{align}
  Further, when $(\gamma_1,\gamma_2)\in\cup_{k=1,2,4}\mathbf R_k$, $\gamma_2\leq\gamma_1<\frac{\alpha}{1-\alpha}$ holds, based on which we have
  \begin{align}\label{Ineq_phii11_phii22}
  \left\{
  \begin{array}{ll}
  \phi_1^{2,2}>\phi_1^{1,1}>0 , &\text{if}~ \gamma_1<\frac{\alpha}{1-\alpha}-1, \\
  \phi_1^{1,1}\geq \phi_1^{2,2}>0, &\text{otherwise},
  \end{array}
  \right.
  ~\text{and}~
  \phi_2^{1,1}\geq \phi_2^{2,2}>0.
  \end{align}
  Applying (\ref{Ineq_phii11_phii22}), $\max\{\Phi^{1,1,1},\Phi^{2,2,2}\}$ under $(\gamma_1,\gamma_2)\in\cup_{k=1,2,4}\mathbf R_k$ can be calculated as
  \begin{align}\label{compare_11_22}
  \max\{\Phi^{1,1,1},\Phi^{2,2,2}\}=
  \left\{
  \begin{array}{ll}
  \Phi^{1,1,1}, &\text{if}~\gamma_1<\frac{\alpha}{1-\alpha}-1, \frac{\rho|h_1|^2}{\phi_1^{2,2}}<\frac{\rho|h_2|^2}{\phi_2^{1,1}},\\
  \Phi^{2,2,2}, &\text{if}~\gamma_1<\frac{\alpha}{1-\alpha}-1, \frac{\rho|h_1|^2}{\phi_1^{2,2}}\geq\frac{\rho|h_2|^2}{\phi_2^{1,1}},\\
                  &\text{or}~\gamma_1\geq\frac{\alpha}{1-\alpha}-1.
  \end{array}
  \right.
  \end{align}
  Moreover, when $(\gamma_1,\gamma_2)\in\cup_{k=1,2,3}\mathbf R_k$, $\gamma_2<\frac{1-\alpha}{\alpha}$ holds, based on which we have
  \begin{align}\label{Ineq_phii12_phii22}
  \left\{
  \begin{array}{ll}
  \phi_1^{1,2}\geq \phi_1^{2,2}>0, &\text{if}~ \gamma_2\geq\frac{(1-\alpha)\gamma_1}{1-\alpha+\alpha\gamma_1}, \\
  \phi_1^{2,2}>\phi_1^{1,2}>0, &\text{otherwise},
  \end{array}
  \right.
  ~\text{and}~
  \phi_2^{1,2}\geq \phi_2^{2,2}>0.
  \end{align}
  Applying (\ref{Ineq_phii12_phii22}), $\max\{\Phi^{1,2,2},\Phi^{2,2,2}\}$ under $(\gamma_1,\gamma_2)\in\cup_{k=1,2,3}\mathbf R_k$ can be calculated as
  \begin{align}\label{compare_12_22}
  \max\{\Phi^{1,2,2},\Phi^{2,2,2}\}=
  \left\{
  \begin{array}{ll}
  \Phi^{1,2,2}, &\text{if}~\gamma_2<\frac{(1-\alpha)\gamma_1}{1-\alpha+\alpha\gamma_1}, \frac{\rho|h_1|^2}{\phi_1^{2,2}}<\frac{\rho|h_2|^2}{\phi_2^{1,2}},\\
  \Phi^{2,2,2}, &\text{if}~\gamma_2<\frac{(1-\alpha)\gamma_1}{1-\alpha+\alpha\gamma_1}, \frac{\rho|h_1|^2}{\phi_1^{2,2}}\geq\frac{\rho|h_2|^2}{\phi_2^{1,2}},\\
                  &\text{or}~\gamma_2\geq\frac{(1-\alpha)\gamma_1}{1-\alpha+\alpha\gamma_1}.
  \end{array}
  \right.
  \end{align}
  Then, combining (\ref{compare_11_22}), (\ref{compare_12_22}), $\gamma_2< \frac{(1-\alpha)\gamma_1}{\alpha+(2\alpha-1)\gamma_1}$ for $(\gamma_1,\gamma_2)\in\mathbf R_3$, and $\gamma_2\geq \frac{(1-\alpha)\gamma_1}{\alpha+(2\alpha-1)\gamma_1}$ for $(\gamma_1,\gamma_2)\in\mathbf R_4$ with (\ref{max_Phi_Oprime_1to4}), after some manipulations we obtain
  \begin{align}\label{best_A1}
    \max_{\pi\in\mathbf P}\Phi^{\pi,\omega_1^\prime(\pi),\omega_2^\prime(\pi)}
    =\left\{
    \begin{array}{ll}
    \Phi^{1,1,1}, &\text{if}~(\gamma_1,\gamma_2)\in\mathbf S_2, \frac{|h_1|^2}{|h_2|^2}<\frac{\phi_1^{2,2}}{\phi_2^{1,1}}, \\\relax
    \Phi^{1,2,2}, &\text{if}~(\gamma_1,\gamma_2)\in\mathbf S_3, \frac{|h_1|^2}{|h_2|^2}<\frac{\phi_1^{2,2}}{\phi_2^{1,2}}, \\\relax
    \Phi^{2,2,2}, &\text{otherwise}.
    \end{array}
    \right.
  \end{align}

  On the other hand, when $(\gamma_1,\gamma_2)\in\mathbf R_5$, we have $\mathbf P=\{2\}$, applying which with (\ref{O_iprime_P2}) we obtain $\max_{\pi\in\mathbf P}\Phi^{\pi,\omega_1^\prime(\pi),\omega_2^\prime(\pi)} = \Phi^{2,2,2}$. Meanwhile, $\mathbf R_5\cap (\mathbf S_2\cup\mathbf S_3)=\emptyset$ can be readily verified by using (\ref{Region_Rk}) and (\ref{Def_R_HUF}), which indicates that (\ref{best_A1}) also results in $\max_{\pi\in\mathbf P}\Phi^{\pi,\omega_1^\prime(\pi),\omega_2^\prime(\pi)}= \Phi^{2,2,2}$ when $(\gamma_1,\gamma_2)\in\mathbf R_5$. Namely, (\ref{best_A1}) is valid for $(\gamma_1,\gamma_2)\in\mathbf R^{\text{ava}}$. At last, by combining (\ref{best_A1}) with (\ref{maxmaxPhi_re1}), the proof is completed.

\section{Proof of Proposition \ref{Prop_cond_sys}}\label{Proof_Prop_cond_sys}
  With regard to the first statement in Proposition \ref{Prop_cond_sys}, when $(\gamma_1, \gamma_2)\in\mathbf R_0$, $\mathcal P_{\text{sys}}^{\text{min}}=1$ is readily obtained from (\ref{Psys_re}) and Lemma \ref{lemma_maxPhi}. Meanwhile, when $(\gamma_1, \gamma_2)\in\cup_{k=1,2,3}\mathbf S_k$, the strategy $\mathcal A^*$ with $[\pi^{\mathcal A^*},\omega_1^{\mathcal A^*},\omega_2^{\mathcal A^*}]=[2,2,2]$ must satisfy $\pi^{\mathcal A^*}\in\mathbf P$ and $\omega_1^{\mathcal A^*},\omega_2^{\mathcal A^*}\in \mathbf O(\pi^{\mathcal A^*})$. Applying this fact with (\ref{Psys_re}) and (\ref{Def_Phi}), we have $\mathcal P_{\text{sys}}^{\text{min}}\leq\mathcal P_{\text{sys}}^{\mathcal A^*}<1$.

  With regard to the second statement, when $(\gamma_1, \gamma_2)\in\mathbf S_1$, we have $\max_{\pi,\omega_1,\omega_2} \Phi^{\pi,\omega_1,\omega_2}=\Phi^{2,2,2}$ according to Lemma \ref{lemma_maxPhi}, which leads to $\mathcal P\{\Phi^{\pi^{\mathcal A^*},\omega_1^{\mathcal A^*},\omega_2^{\mathcal A^*}}<1,\max_{\pi,\omega_1,\omega_2\in\{1,2\}}\Phi^{\pi,\omega_1,\omega_2}\geq 1\}=0$. Combining this fact with Lemma \ref{lemma_NScondition}, $\mathcal P_{\text{sys}}^{\mathcal A^*}=\mathcal P_{\text{sys}}^{\text{min}}$ is proved.

  For the third statement, we first focus on the proof in case $(\gamma_1, \gamma_2)\in\mathbf S_2$. Recall that $(\gamma_1, \gamma_2)\in\mathbf S_2$ ensures $\gamma_2\geq \frac{(1-\alpha)\gamma_1}{\alpha+(2\alpha-1)\gamma_1}$, applying which with (\ref{O_iprime_P1}) and (\ref{O_iprime_P2}) we obtain $\Phi^{1,1,1}\geq\Phi^{1,\omega_1,\omega_2}$ for any $\omega_1,\omega_2\in\mathbf O(1)$, and $\Phi^{2,2,2}\geq\Phi^{2,\omega_1,\omega_2}$ for any $\omega_1,\omega_2\in\mathbf O(2)$. Combining this result with (\ref{Def_Phi}) we have $\mathcal P\{\Phi^{1,1,1}<1\}\leq\mathcal P\{\Phi^{1,\omega_1,\omega_2}<1\}$ and $\mathcal P\{\Phi^{2,2,2}<1\}\leq\mathcal P\{\Phi^{2,\omega_1,\omega_2}<1\}$ for any $\omega_1,\omega_2\in\{1,2\}$.
  This fact reveals that, to prove there is no channel-unrelated strategy achieving $\mathcal P_{\text{sys}}^{\text{min}}$, we only need to prove channel unrelated decisions $[\pi,\omega_1,\omega_2]=[1,1,1]$ and $[\pi,\omega_1,\omega_2]=[2,2,2]$ cannot achieve $\mathcal P_{\text{sys}}^{\text{min}}$. To this end, we construct a channel-related strategy $\mathcal A^\dagger$ as $[\pi^{\mathcal A^\dagger},\omega_1^{\mathcal A^\dagger},\omega_2^{\mathcal A^\dagger}]=[1,1,1]$ if ${|h_1|^2}/{|h_2|^2}<{\phi_1^{2,2}}/{\phi_2^{1,1}}$, and $[\pi^{\mathcal A^\dagger},\omega_1^{\mathcal A^\dagger},\omega_2^{\mathcal A^\dagger}]=[2,2,2]$ otherwise.
  Applying $\mathcal A^\dagger$, we have
  \begin{align}\label{P1_neq_0}
    &\mathcal P\{\Phi^{1,1,1}<1,\Phi^{\pi^{\mathcal A^\dagger},\omega_1^{\mathcal A^\dagger},\omega_2^{\mathcal A^\dagger}}\geq 1\}
    =\mathcal P \{ \Phi^{1,1,1}<1,\Phi^{2,2,2}\geq 1, \tfrac{|h_1|^2}{\phi_1^{2,2}}\geq\tfrac{|h_2|^2}{\phi_2^{1,1}}\}\notag\\
    =&\mathcal P \{ {\phi_2^{1,1}}>{\rho|h_2|^2}\geq{\phi_2^{2,2}}\}
      \mathcal P \{ {\rho|h_1|^2}\geq {\phi_1^{2,2}}\}
    >0,
  \end{align}
  where the second equality is obtained by substituting (\ref{Def_Phi}) and applying $\phi_2^{1,1}> \phi_2^{2,2}>0$ when $(\gamma_1, \gamma_2)\in\mathbf S_2$.
  Similarly, by employing $\mathcal A^\dagger$ and $\phi_1^{2,2}> \phi_1^{1,1}>0$ for $(\gamma_1, \gamma_2)\in\mathbf S_2$, we also have
  \begin{align}\label{P2_neq_0}
    \mathcal P\{\Phi^{2,2,2}<1,\Phi^{\pi^{\mathcal A^\dagger},\omega_1^{\mathcal A^\dagger},\omega_2^{\mathcal A^\dagger}}\geq 1\}
    =\mathcal P \{ {\phi_1^{2,2}}>{\rho|h_1|^2}\geq{\phi_1^{1,1}}\}
      \mathcal P \{ {\rho|h_2|^2}\geq {\phi_2^{1,1}}\}
    >0.
  \end{align}
  Combining the above facts with Lemma \ref{lemma_NScondition}, it is proved that channel-unrelated strategies cannot achieve $\mathcal P_{\text{sys}}^{\text{min}}$ when $(\gamma_1, \gamma_2)\in\mathbf S_2$. Following the same rationale, it can be proved that channel-unrelated strategies cannot achieve $\mathcal P_{\text{sys}}^{\text{min}}$ when $(\gamma_1, \gamma_2)\in\mathbf S_3$. This completes the proof.

\section{Proof of Proposition \ref{Prop_cond_user}}\label{Proof_Prop_cond_user}
The first two statements of Proposition \ref{Prop_cond_user} can be straightly obtained from Lemma \ref{Lemma_Pimin}, and thus we focus on the proof of the third statement. Considering $(\gamma_1, \gamma_2)\in\mathbf S_2$, we assume that there exists a strategy $\mathcal A$ which achieves $\mathcal P^{\mathcal A}_i=\mathcal P^{\text{min}}_i$ for $i=1,2$, simultaneously. According to (\ref{Form_Pi}) and (\ref{Event_Si}), $\mathcal A$ should satisfy $\pi^{\mathcal A}\in\mathbf P$ and $\omega_i^{\mathcal A}\in\mathbf O(\pi^{\mathcal A})$, and thus we have
\begin{align}\label{P_1A}
  \mathcal P^{\mathcal A}_1
  =\mathcal P\{\rho|h_1|^2<\phi_1^{\pi^{\mathcal A},\omega_1^{\mathcal A}}\}
  =\mathcal P\{\rho|h_1|^2<\phi^{1,1}_1\}+\mathcal P\{\rho|h_1|^2<\phi_1^{\pi^{\mathcal A},\omega_1^{\mathcal A}},\rho|h_1|^2\geq\phi^{1,1}_1\},
\end{align}
where the second equality is obtained by using the law of total probability and $\phi^{1,1}_1<\phi^{\pi,\omega_1}_1$ for any $\pi\in\mathbf P$ and $\omega_1\in\mathbf O(\pi)$ when $(\gamma_1, \gamma_2)\in\mathbf S_2$. Using (\ref{P_1A}) and $\mathcal P^{\text{min}}_1=\mathcal P\{\rho|h_1|^2<\phi^{1,1}_1\}$ given by Lemma \ref{Lemma_Pimin}, $\mathcal P^{\mathcal A}_1=\mathcal P^{\text{min}}_1$ leads to
\begin{align}\label{condiPr1_1}
  &\mathcal P\{\phi_1^{\pi^{\mathcal A},\omega_1^{\mathcal A}}>\rho|h_1|^2\geq\phi^{1,1}_1\}=0 \notag\\
  =&
  \sum_{\pi\in\mathbf P,\omega_1\in\mathbf O(\pi)}
  \mathcal P\{\phi^{\pi,\omega_1}_1>\rho|h_1|^2\geq\phi^{1,1}_1, (\pi^{\mathcal A},\omega_1^{\mathcal A})=(\pi,\omega_1)\} \notag\\
  =&
  \sum_{\pi\in\mathbf P,\omega_1\in\mathbf O(\pi),(\pi,\omega_1)\neq(1,1)}
  \mathcal P\{\phi^{\pi,\omega_1}_1>\rho|h_1|^2\geq\phi^{1,1}_1, (\pi^{\mathcal A},\omega_1^{\mathcal A})=(\pi,\omega_1)\}.
\end{align}
Further, since when $(\gamma_1, \gamma_2)\in\mathbf S_2$, $\phi_1^{2,2}\leq\phi_1^{\pi,\omega_1}$ holds for any $\pi$ and $\omega_1$ satisfying $\pi\in\mathbf P$, $\omega_1\in\mathbf O(\pi)$, and $(\pi,\omega_1)\neq(1,1)$, (\ref{condiPr1_1}) leads to
\begin{align}\label{condiPr1}
  &\sum_{\pi\in\mathbf P,\omega_1\in\mathbf O(\pi),(\pi,\omega_1)\neq(1,1)}
  \mathcal P\{\phi^{2,2}_1>\rho|h_1|^2\geq\phi^{1,1}_1, (\pi^{\mathcal A},\omega_1^{\mathcal A})=(\pi,\omega_1)\}=0  \notag\\
  \Leftrightarrow&
  \mathcal P\Big\{\underset{\pi\in\mathbf P,\omega_1\in\mathbf O(\pi),(\pi,\omega_1)\neq(1,1)}{\cup} \big(\phi^{2,2}_1>\rho|h_1|^2\geq\phi^{1,1}_1, (\pi^{\mathcal A},\omega_1^{\mathcal A})=(\pi,\omega_1) \big) \Big\}=0 \notag\\
  \Leftrightarrow&
  \mathcal P\Big\{\phi^{2,2}_1>\rho|h_1|^2\geq\phi^{1,1}_1,
  \underset{\pi\in\mathbf P,\omega_1\in\mathbf O(\pi),(\pi,\omega_1)\neq(1,1)}{\cup}  (\pi^{\mathcal A},\omega_1^{\mathcal A})=(\pi,\omega_1) \Big\}=0 \notag\\
  \Leftrightarrow&
  \mathcal P\Big\{ \underset{\pi\in\mathbf P,\omega_1\in\mathbf O(\pi),(\pi,\omega_1)\neq(1,1)}{\cup}  (\pi^{\mathcal A},\omega_1^{\mathcal A})=(\pi,\omega_1)
  \big|  \phi^{2,2}_1>\rho|h_1|^2\geq\phi^{1,1}_1 \Big\}=0 \notag\\
  \Leftrightarrow&
  \mathcal P\{ (\pi^{\mathcal A},\omega_1^{\mathcal A})=(1,1)
  \big|  \phi^{2,2}_1>\rho|h_1|^2\geq\phi^{1,1}_1 \}=1.
\end{align}
On the other hand, by following the similar steps shown above, $\mathcal P^{\mathcal A}_2=\mathcal P^{\text{min}}_2$ results in
\begin{align}\label{condiPr2}
  \mathcal P\{ (\pi^{\mathcal A},\omega_2^{\mathcal A})=(2,2)
  \big|  \phi_2^{1,1}>\rho|h_2|^2\geq\phi_2^{2,2} \}=1.
\end{align}
Recall that, $\mathcal P\{\mathbb A_1 | \mathbb B_1\}=\mathcal P\{\mathbb A_2 | \mathbb B_2\}=1$ indicates $\mathbb A_1\supseteq \mathbb B_1$ and $\mathbb A_2\supseteq \mathbb B_2$, which leads to $\mathcal P\{\mathbb A_1,\mathbb A_2 , \mathbb B_1, \mathbb B_2\}=\mathcal P\{\mathbb B_1, \mathbb B_2\}$, where $\mathbb A_i$ and $\mathbb B_i$ ($i=1,2$) denote arbitrary events. Employing this fact and $\mathcal P\{\mathbb A_1,\mathbb A_2\}\geq\mathcal P\{\mathbb A_1,\mathbb A_2 , \mathbb B_1, \mathbb B_2\}$, (\ref{condiPr1}) and (\ref{condiPr2}) lead to $\mathcal P\{ (\pi^{\mathcal A},\omega_1^{\mathcal A})=(1,1), (\pi^{\mathcal A},\omega_2^{\mathcal A})=(2,2)\}\geq
\mathcal P\{\phi^{2,2}_1>\rho|h_1|^2\geq\phi^{1,1}_1\}\mathcal P\{ \phi_2^{1,1}>\rho|h_2|^2\geq\phi_2^{2,2}\}>0$. This inequality indicates that, strategy $\mathcal A$ simultaneously selects $\pi^{\mathcal A}=1$ and $\pi^{\mathcal A}=2$ with a nonzero probability, which makes $\mathcal A$ nonexistent. Following the above steps, the same result can be proved for $(\gamma_1, \gamma_2)\in\mathbf S_3$, which completes the proof of Proposition \ref{Prop_cond_user}.

\section{Proof of Corollary \ref{Coro_HUF}}\label{Proof_Coro_HUF}
We first prove $\mathcal P^{\mathcal H}_{\text{sys}}=\mathcal P^{\text{min}}_{\text{sys}}$ when $(\gamma_1,\gamma_2)\in\mathbf R^{\text{ava}}$. According to the proposed HUF strategy and Proposition \ref{Prop_cond_sys}, $\mathcal P^{\mathcal H}_{\text{sys}}=\mathcal P^{\text{min}}_{\text{sys}}$ for $(\gamma_1,\gamma_2)\in\mathbf S_1$ is directly obtained. When $(\gamma_1,\gamma_2)\in\mathbf S_2$, by applying the HUF strategy, Lemma \ref{lemma_maxPhi}, and the law of total probability, we have
\begin{align}\label{Preq0}
   &\mathcal P\{ \Phi^{\pi^{\mathcal H},\omega_1^{\mathcal H},\omega_2^{\mathcal H}}<1,\max_{\pi,\omega_1,\omega_2\in\{1,2\}} \Phi^{\pi,\omega_1,\omega_2}\geq 1 \}\notag\\
  =&\mathcal P\{ \Phi^{2,2,2}<1, \Phi^{1,1,1}\geq 1, \tfrac{\rho|h_2|^2}{\phi_2^{1,1}}>\tfrac{\rho|h_1|^2}{\phi_1^{2,2}}>1 \}
    +\mathcal P\{ \Phi^{1,1,1}<1, \Phi^{2,2,2}\geq 1, \tfrac{\rho|h_2|^2}{\phi_2^{1,1}} \leq \tfrac{\rho|h_1|^2}{\phi_1^{2,2}} \leq 1\}\notag\\
  =&\mathcal P\{ \tfrac{\rho|h_1|^2}{\phi_1^{2,2}}<1, \tfrac{\rho|h_2|^2}{\phi_2^{1,1}}>\tfrac{\rho|h_1|^2}{\phi_1^{2,2}}>1 , \Phi^{1,1,1}\geq 1\}
    +\mathcal P\{\tfrac{\rho|h_1|^2}{\phi_1^{2,2}}=1, \tfrac{\rho|h_2|^2}{\phi_2^{2,2}}\geq 1, \tfrac{\rho|h_2|^2}{\phi_2^{1,1}}\leq 1, \Phi^{1,1,1}<1\}\notag\\
  =&0  ,
\end{align}
where the second equality is obtained by substituting $\Phi^{2,2,2}$ given by (\ref{Def_Phi}) and applying $\phi_2^{2,2}<\phi_2^{1,1}$ for $(\gamma_1,\gamma_2)\in\mathbf S_2$. Combining (\ref{Preq0}) and Lemma \ref{lemma_NScondition}, $\mathcal P^{\mathcal H}_{\text{sys}}=\mathcal P^{\text{min}}_{\text{sys}}$ for $(\gamma_1,\gamma_2)\in\mathbf S_2$ is proved. Following the same rationale, $\mathcal P^{\mathcal H}_{\text{sys}}=\mathcal P^{\text{min}}_{\text{sys}}$ for $(\gamma_1,\gamma_2)\in\mathbf S_3$ can be proved.

On the other hand, when $(\gamma_1,\gamma_2)\in\mathbf S_1$, $\mathcal P^{\mathcal H}_i=\mathcal P^{\text{min}}_i$ for $i=1,2$ can be straightly obtained by using Proposition \ref{Prop_cond_user}. When $(\gamma_1,\gamma_2)\in\mathbf S_2$, according to (\ref{Form_Pi}) and (\ref{Event_Si}) we have
\begin{align}\label{Phuf1_eq_min}
  \mathcal P^{\mathcal H}_1
  =&\mathcal P\{ \rho|h_1|^2<\phi_1^{2,2}, \rho|h_1|^2\geq\phi_1^{2,2} \}+\mathcal P\{ \rho|h_1|^2<\phi_1^{1,1}, \rho|h_1|^2<\phi_1^{2,2} \}\notag\\
  =&\mathcal P\{ \rho|h_1|^2<\phi_1^{1,1} \},
\end{align}
where the second equality is guaranteed by $\phi_1^{1,1}<\phi_1^{2,2}$ when $(\gamma_1,\gamma_2)\in\mathbf S_2$. Combining (\ref{Phuf1_eq_min}) with Lemma \ref{Lemma_Pimin}, $\mathcal P^{\mathcal H}_1=\mathcal P^{\text{min}}_1$ is proved for $(\gamma_1,\gamma_2)\in\mathbf S_2$. Following the same rationale, $\mathcal P^{\mathcal H}_1=\mathcal P^{\text{min}}_1$ for $(\gamma_1,\gamma_2)\in\mathbf S_3$ can be obtained. This completes the proof.



\begin{thebibliography}{99}

\bibitem{Yuan_survey}
Y. Yuan et al., ``NOMA for next-generation massive IoT: Performance potential and technology directions,'' \emph{IEEE Commun. Mag.}, vol. 59, no. 7, pp. 115-121, Jul. 2021.


\bibitem{Islam_survey}
S. M. R. Islam, N. Avazov, O. A. Dobre, and K. Kwak, ``Power-domain non-orthogonal multiple access (NOMA) in 5G systems: Potentials and challenges,'' \emph{IEEE Commun. Surveys Tuts.}, vol. 19, no. 2, pp. 721-742, 2nd Quart., 2017.


\bibitem{NZhao_TCOM}
N. Zhao et al., ``Joint beamforming and jamming optimization for secure transmission in MISO-NOMA networks,'' \emph{IEEE Trans. Commun.}, vol. 67, no. 3, pp. 2294-2305, Mar. 2019.

\bibitem{YXu_adaptive}
Y. Xu, J. Tang, B. Li, N. Zhao, D. Niyato, and K. -K. Wong, ``Adaptive aggregate transmission for device-to-multi-device aided cooperative NOMA networks,'' \emph{IEEE J. Sel. Areas Commun.}, vol. 40, no. 4, pp. 1355-1370, Apr. 2022.

\bibitem{YXu_robust}
Y. Xu, R. Q. Hu, and G. Li, ``Robust energy-efficient maximization for cognitive NOMA networks under channel uncertainties,'' \emph{IEEE Internet Things J.}, vol. 7, no. 9, pp. 8318-8330, Sep. 2020.


\bibitem{Ycao_TWC}
Y. Cao et al., ``Privacy preservation via beamforming for NOMA,'' \emph{IEEE Trans. Wireless Commun.}, vol. 18, no. 7, pp. 3599-3612, Jul. 2019.

\bibitem{HLei_TCOM}
H. Lei, R. Gao, K. -H. Park, I. S. Ansari, K. J. Kim, and M. -S. Alouini, ``On secure downlink NOMA systems with outage constraint,'' \emph{IEEE Trans. Commun.}, vol. 68, no. 12, pp. 7824-7836, Dec. 2020.

\bibitem{Ding_impact}
Z. Ding, P. Fan, and H. V. Poor, ``Impact of user pairing on 5G nonorthogonal multiple-access downlink transmissions,'' \emph{IEEE Trans. Veh. Technol.}, vol. 65, no. 8, pp. 6010-6023, Aug. 2016.


\bibitem{FZhou_CRNOMA}
F. Zhou, Y. Wu, Y. Liang, Z. Li, Y. Wang, and K. Wong, ``State of the art, taxonomy, and open issues on cognitive radio networks with NOMA,'' \emph{IEEE Wireless Commun.}, vol. 25, no. 2, pp. 100-108, Apr. 2018.

\bibitem{BChen_CRNOMA}
B. Chen, Y. Chen, Y. Chen, Y. Cao, N. Zhao, and Z. Ding, ``A novel spectrum sharing scheme assisted by secondary NOMA relay,'' \emph{IEEE Wireless Commun. Lett.}, vol. 7, no. 5, pp. 732-735, Oct. 2018.

\bibitem{ZYang_general}
Z. Yang, Z. Ding, P. Fan, and N. Al-Dhahir, ``A general power allocation scheme to guarantee quality of service in downlink and uplink NOMA systems,'' \emph{IEEE Trans. Wireless Commun.}, vol. 15, no. 11, pp. 7244-7257, Nov. 2016.


\bibitem{ZXiao_fairness}
Z. Xiao, L. Zhu, Z. Gao, D. O. Wu, and X. -G. Xia, ``User fairness non-orthogonal multiple access (NOMA) for millimeter-wave communications with analog beamforming,'' \emph{IEEE Trans. Wireless Commun.}, vol. 18, no. 7, pp. 3411-3423, Jul. 2019.

\bibitem{Ding_random_users}
Z. Ding, Z. Yang, P. Fan, and H. V. Poor, ``On the performance of non-orthogonal multiple access in 5G systems with randomly deployed users,'' \emph{IEEE Signal Process. Lett.}, vol. 21, no. 12, pp. 1501-1505, Dec. 2014.


\bibitem{THou_nakagami}
T. Hou, X. Sun, and Z. Song, ``Outage performance for non-orthogonal multiple access with fixed power allocation over Nakagami-${m}$ fading channels,'' \emph{IEEE Commun. Lett.}, vol. 22, no. 4, pp. 744-747, Apr. 2018.

\bibitem{ZYang_imperfectCSI}
Z. Yang, Z. Ding, P. Fan, and G. K. Karagiannidis, ``On the performance of non-orthogonal multiple access systems with partial channel information,'' \emph{IEEE Trans. Commun.}, vol. 64, no. 2, pp. 654-667, Feb. 2016.

\bibitem{Ionnis_fairness}
S. Timotheou and I. Krikidis, ``Fairness for non-orthogonal multiple access in 5G systems,'' \emph{IEEE Signal Process. Lett.}, vol. 22, no. 10, pp. 1647-1651, Oct. 2015.

\bibitem{GLi_RA}
G. Li, D. Mishra, and H. Jiang, ``Resource allocation in power-beacon-assisted IoT networks with nonorthogonal multiple access,'' \emph{IEEE Internet Things J.}, vol. 8, no. 18, pp. 14385-14398, Sep. 2021.


\bibitem{Sha_grantfree}
M. B. Shahab, R. Abbas, M. Shirvanimoghaddam, and S. J. Johnson, ``Grant-free non-orthogonal multiple access for IoT: A survey,'' \emph{IEEE Commun. Surveys Tuts.}, vol. 22, no. 3, pp. 1805-1838, 3rd Quart., 2020.

\bibitem{XLi_B5G}
X. Li et al., ``Cooperative wireless-powered NOMA relaying for B5G IoT networks with hardware impairments and channel estimation errors,'' \emph{IEEE Internet Things J.}, vol. 8, no. 7, pp. 5453-5467, Apr. 2021.

\bibitem{PXu_1bit}
P. Xu, Y. Yuan, Z. Ding, X. Dai, and R. Schober, ``On the outage performance of non-orthogonal multiple access with 1-bit feedback,'' \emph{IEEE Trans. Wireless Commun.}, vol. 15, no. 10, pp. 6716-6730, Oct. 2016.

\bibitem{Kim_CMD}
J. Kim and I. Lee, ``Capacity analysis of cooperative relaying systems using non-orthogonal multiple access,'' \emph{IEEE Commun. Lett.}, vol. 19, no. 11, pp. 1949-1952, Nov. 2015.

\bibitem{Gong_CMD}
M. Gong and Z. Yang, ``The application of antenna diversity to NOMA with statistical channel state information,'' \emph{IEEE Trans. Veh. Technol.}, vol. 68, no. 4, pp. 3755-3765, Apr. 2019.

\bibitem{YLiu_CoNOMA}
Y. Liu, Z. Ding, M. Elkashlan, and H. V. Poor, ``Cooperative non-orthogonal multiple access with simultaneous wireless information and power transfer,'' \emph{IEEE J. Sel. Areas Commun.}, vol. 34, no. 4, pp. 938-953, Apr. 2016.

\bibitem{HWang_HARQ}
H. Wang, Z. Shi, Y. Fu, and R. Song, ``Outage performance for NOMA-aided small cell networks with HARQ,'' \emph{IEEE Wireless Commun. Lett.}, vol. 10, no. 1, pp. 72-76, Jan. 2021.




\bibitem{YXu_TWC}
Y. Xu, J. Cheng, G. Wang, and V. C. M. Leung, ``Adaptive coordinated direct and relay transmission for NOMA networks: A joint downlink-uplink scheme,'' \emph{IEEE Trans. Wireless Commun.}, vol. 20, no. 7, pp. 4328-4346, Jul. 2021.

\bibitem{GX_DD}
G. Li, D. Mishra, and H. Jiang, ``Cooperative NOMA with incremental relaying: Performance analysis and optimization,'' \emph{IEEE Trans. Veh. Technol.}, vol. 67, no. 11, pp. 11291-11295, Nov. 2018.

\bibitem{Ding_QoS}
Z. Ding, H. Dai, and H. V. Poor, ``Relay selection for cooperative NOMA,'' \emph{IEEE Wireless Commun. Lett.}, vol. 5, no. 4, pp. 416-419, Aug. 2016.

\bibitem{Lulv_QoS}
L. Lv, J. Chen, Q. Ni, and Z. Ding, ``Design of cooperative non-orthogonal multicast cognitive multiple access for 5G systems: User scheduling and performance analysis,'' \emph{IEEE Trans. Commun.}, vol. 65, no. 6, pp. 2641-2656, Jun. 2017.

\bibitem{LY_QoS}
L. Yang et al., ``Cooperative non-orthogonal layered multicast multiple access for heterogeneous networks,'' \emph{IEEE Trans. Commun.}, vol. 67, no. 2, pp. 1148-1165, Feb. 2019.

\bibitem{KSAli}
K. S. Ali, M. Haenggi, H. ElSawy, A. Chaaban, and M. -S. Alouini, ``Downlink non-orthogonal multiple access (NOMA) in Poisson networks,'' \emph{IEEE Trans. Commun.}, vol. 67, no. 2, pp. 1613-1628, Feb. 2019.

\bibitem{MUST}
``Study on downlink multiuser superposition transmission (MUST) for LTE,'' 3GPP, Tech. Rep. TR 36.859.

\bibitem{Ding_unveiling}
Z. Ding, R. Schober, and H. V. Poor, ``Unveiling the importance of SIC in NOMA systems-Part 1: State of the art and recent findings,'' \emph{IEEE Commun. Lett.}, vol. 24, no. 11, pp. 2373-2377, Nov. 2020.

\bibitem{Common_myth}
M. Vaezi, R. Schober, Z. Ding, and H. V. Poor, ``Non-orthogonal multiple access: Common myths and critical questions,'' \emph{IEEE Wireless Commun.}, vol. 26, no. 5, pp. 174-180, Oct. 2019.


\end{thebibliography}
\end{document}